\documentclass{aa} 
\usepackage{graphicx}
\usepackage{txfonts}
\usepackage{verbatim}
\usepackage{natbib}
\usepackage{multirow}

\usepackage{color}

\bibpunct{(}{)}{;}{a}{}{,} 

\usepackage[]{hyperref}

\newcommand{\iti}[1]{\textit{\textbf{#1}}}
\newcommand{\jpgeps}{jpg}

\begin{document}

\title{Method of frequency dependent correlations: investigating the variability of total solar irradiance}
\titlerunning {Method of frequency dependent correlations} 

\author{J. Pelt \inst{1} \and
           M.~J. K\"apyl\"a  \inst{2,3} \and
           N. Olspert \inst{2}}

\institute{Tartu Observatory, T\~{o}ravere, 61602, Estonia
                 \and  
                 ReSoLVE Centre of Excellence, Department of Computer
                 Science, Aalto University, PO Box 15400, FI-00076 Aalto, Finland
                 \and
                 Max Planck Institute for Solar System Research, Justus-von-Liebig-Weg 3, D-37077 G\"ottingen, Germany
          }

\offprints{J. Pelt\\
          \email{pelt@aai.ee}
          }   

\date{Received ; accepted}

\abstract
{
This paper contributes to the field of modeling and hindcasting of the total solar
   irradiance (TSI) based on different proxy data that extend
further    back in time than the TSI that is measured from satellites.
}
{We introduce a simple method to analyze persistent
  frequency-dependent correlations (FDCs) between the time series and
  use these correlations to hindcast missing historical TSI values.
  We try to avoid arbitrary choices of the free parameters of the model by computing them using an optimization procedure. 
  The method can be regarded as a general tool for pairs of data sets, where correlating and anticorrelating components 
  can be separated into non-overlapping regions in frequency domain.}  
{Our method is based on low-pass and band-pass filtering with a
  Gaussian transfer function combined with de-trending and computation
  of envelope curves.}
{
We find a major controversy between the historical proxies and satellite-measured targets: a large
  variance is detected between the low-frequency parts of
  targets, while the low-frequency proxy behavior of different measurement series is consistent with high
  precision. We also show that even though the rotational signal is
  not strongly manifested in the targets and proxies, it becomes
  clearly visible in FDC spectrum.  A significant part of the variability
  can be explained by a very simple model consisting of
  two components: the original proxy describing blanketing by sunspots, and
  the low-pass-filtered curve describing the overall
  activity level.
  The models with the full library of the different building
  blocks can be applied to hindcasting with a high level of confidence, $R_c\approx 0.90$. 
  The usefulness of these models is limited by the major target controversy.
}
{The application of the new method to solar data allows us to obtain important insights into the different TSI modeling procedures and 
  their capabilities for hindcasting based on the directly observed time intervals.}


\maketitle

\section{Introduction}\label{s:intro}

The total solar irradiance (TSI) has only been directly measured
since 1978, and the available data roughly cover three and half solar cycles. From these
measurements it is evident \citep[see, e.g.,][and references
therein]{Froehlich13} that on average, the maximum to minimum
variation during the solar cycle is roughly 0.1\%. The measurements
also show the last
prolonged minimum marking the transition between cycles 23 and 24 has
been unusual, with very low activity accompanied with an
extremely low TSI. The TSI value measured during the 
last solar minimum, at the end of 2008, was
significantly lower than the TSI of the two previous minima. This has been
postulated to be indicative of a long-term decreasing trend since 1985
\citep{LF07,LF08}. This finding might have major implications for the studies
of climate and global warming on Earth, but also for the solar
physics community, because the observed major change in the overall solar
activity level might mark a disruption of the dynamo process
that generates the solar magnetic field.
The time range of the
direct TSI measurements is far too short to estimate whether there is such a trend, and if it is there, how
significant it is on longer timescales.
Proxies or extrapolation-based ways of reconstructing the longer term evolution are therefore required.

There are several ways of reconstructing the TSI back in time
that vary in the level of complication and time extent from purely empirical to
physically motivated models that use several constituents that affect
TSI. 
The longest reconstructions of up to 10 000 years and even
longer back in time
can be obtained using cosmogenic isotopes \citep[see, e.g.,][]{Steinhilber09,Vieira2011}, while sunspot number
and area recordings provide a time window of roughly 300-400 years
\citep[see, e.g.,][]{SolankiKrivova04}. The geomagnetic AA index has
also been used as a proxy for reconstructions of the TSI until the late nineteenth
century \citep[see, e.g.,][]{Rouillard07}. 

At the simplest level, the models are based on linear, nonlinear, or
multivariate regressions of some set of proxy variables \citep[see,
e.g.,][]{FoukalLean88,Chapman96,FroehlichLean98,Fligge98,Preminger02,JainHasan04,Froehlich13}. Physics-based
approaches in general analyze maps of a given proxy that are
transformed to produce irradiance through a process that 
can involve multiple steps and  
model atmospheres \citep[see,
e.g.,][]{Fontenla04,Ermolli03,KrivovaSolanki07}. The 
models typically employ three to seven different components describing
the quiet Sun, sunspot darkening, and brightening by faculae and
network, most often relying on the assumption that the TSI variation is
entirely caused by the magnetic field at the solar surface \citep[see,
e.g.,][and references therein]{KrivovaSolanki07}. 
Some other authors emphasized the effect of
magnetic activity, which produces a global modulation of thermal structure
\citep[see, e.g.,][and references therein]{Li03}. 
Most reconstructions work well on shorter timescales, while the secular change on centennial timescales and longer still remains an open issue.

In this paper, we formulate a new simple method of frequency-dependent correlations (hereafter FDCs) to describe the correlations
of different proxies and the direct (mostly) satellite-based measurements (in the
form of different composites). We show that by using simple devices of statistical signal processing, we can
obtain insights into various problems that occur when we work with
modeling, predicting, and hindcasting of the TSI records.

Among the effects studied are the separation of the proxies and
targets into low-frequency (LF) and high-frequency (HF) components
with low-pass filtering (smoothing). The LF component describes
the smooth cycle behavior, while the HF component characterizes the sharp
dimmings and brightenings caused by the passage of active regions. The
other simple method of computing the correlation spectra using a Gaussian bandpass
filter is used here to study the somewhat paradoxical feature of the
solar rotation being hidden in the raw target and proxy
data.

From the very beginning, we must stress that we place the main
emphasis on the stationary features of the observed time series. The
transients and secular trends then reveal themselves as fitting
residuals.  We try to avoid overparametrization and overfitting, which
occur when there is a desire to minimize these residuals. Small
modeling residuals do not always
mean that the predictive power of the model is good.

The other important aspect of our paper is the simple nature of
the proposed algorithms. 
We are well aware that in principle, more
precise proxy-to-target fitting results can be obtained by very
complex physical modeling of the TSI variability on different timescales. Unfortunately, direct observations of the solar surface are only available for recent
years, which means that they cannot be used for hindcasting past TSI values.  
As we aim to show in this paper, significant insights can
be obtained by using almost trivial methods. The simplest models presented here can be considered to outperform more
complex analyses because they are more transparent, easier to use, and can be more easily
repeated.

The paper is organized as follows.
After introducing all the data sets in Sect.~\ref{s:data}, we
cover the elements of our method in
Sect.~\ref{s:method}. This part can have many more applications than
the fields we investigate here.  Then we apply our methods to
a wide set of well-known proxies and prediction targets. We start
with some simple diagnostic tests (Sect.~\ref{s:diagnostics}) that help
to locate specific problems that are encountered when modeling.
Then we describe almost trivial modeling schemes (Sect.~\ref{s:simple}) 
and introduce more complex solutions
later (Sect.~\ref{s:multimod}). The specific results of our analyses are presented in
Sect.~\ref{ss:compmod}. In the discussion part we place our computed
examples into physical context. Even after quite complex modeling
efforts, some of the variance in the TSI remains unexplained by the
proxies. We link this in Sect.~\ref{s:disc} to possible secular
changes that are most likely related to the modulation of the irradiance through the
changing level of magnetic activity and to still-hidden minor
nonlinearities. Most importantly, however, we discuss the
problem of hindcasting from a somewhat unusual point of view to form an idea about the level of prediction precision that
is achievable
using only simplest devices. We also determine the main obstacles of proper
day-to-day precision TSI estimation.
In Sect.~\ref{s:concl} we present our conclusions.

\section{Data}\label{s:data}

To build models for the targets based on proxies, we
use some standard well-known data sets that are listed in
Table~\ref{table001}. We briefly list the most
important properties of these data sets below for this particular work.

\begin{table*}[!ht]
\caption{Data sets.}
\begin{center}
  \begin{tabular}{lllll}
\hline    
Proxy data set &Start &End &$N_{obs}$ &References \\
\hline
PSI  &09.05.1874 &31.05.2013 &49657 &\cite{Balmaceda2009}$^{1}$ \\
SA   &01.05.1874 &09.09.2014   &51135 &NASA/Marshall Space Flight Center\\
     &           &             &      &Solar Physics web pages$^{2}$\\
SN   &09.01.1818 &30.09.2013   &68249 &World Data Center SILSO,\\
&           &             &      &Royal Observatory of Belgium, Brussels$^{3}$\\
RADIO &14.02.1947 &31.05.2013    &23572 &Laboratory for Atmospheric and Space\\
&             &      &&Physics (Univ. of Colorado, Boulder\\
&             &      &&Time Series Server$^{4}$\\
&             &      &&\cite{Tapping2013}\\
MGII &07.11.1978    &19.03.2015 &13282 &The Global Ozone Monitoring Experiment,\\
&             &      &&Institute of Environmental Physics, Univ. of Bremen$^{5}$\\
&             &      &&\cite{Viereck2004}\\
LYMAN &14.02.1947 &16.03.2015 &24868 &LASP Interactive Solar Irradiance Data Center $^{6}$\\
&             &      &&\cite{Lindholm2011}\\
\hline
Target (TSI) data set      &&&& \\
\hline
ACRIM         &17.11.1978 &17.09.2013 &12158 &Website of ACRIM missions$^{8}$\\
&&&&\cite{Willson2014}\\
&&&&File identifier {\tt acrim\_composite\_131130\_hdr.txt}\\
PMOD          & 17.11.1978  & 03.08.2016 &  13079 &Davos Physical-Meteorological Observatory$^{9}$\\
&&&&File identifier {\tt composite\_42\_65\_1608.dat}\\
RMIB          &02.07.1981 &13.01.2015 &11988 &Royal Meteorological Institute of Belgium$^{10}$\\
\hline
\end{tabular}

\label{table001}
\end{center}
\end{table*}

\subsection{Proxies}\label{ss:proxies}

The first data set we used as a proxy is the Photometric Sunspot Index
(hereafter PSI), which is calculated after cross-calibration of measurements made by
different observatories\footnote{\url{http://www2.mps.mpg.de/projects/sun-climate/data.html}} \citep[see][and references therein]{Balmaceda2009}. 
In the original PSI data, we corrected four
probably outlying observations using tabular interpolated values
instead (specifically for days 05.02.1989, 18.11.1991, 08.10.2000, and
17.06.2011). The results below are practically independent of the
outlier values, but we removed them from the data in any case to
facilitate the plotting of the data. For uniformity with the standard
sunspot area data, we used the original tabulated
PSI values in our computations with reverted sign. In this way, the total sunspot area data
and PSI will correlate positively. 

A data set of sunspot areas
(hereafter SA), as compiled by D. Hathaway and reported by the NASA/Marshall Space Flight Center\footnote{\url{http://solarscience.msfc.nasa.gov/greenwch/daily_area.txt}} 
, is our second proxy. 
We note that another SA compilation by \citet{Balmaceda2009} exists, 
but here we have chosen to use the data compiled by Hathaway. 
Even though the data sets are rather similar, this might explain part of the differences that we see when using PSI and SA.

The third proxy we used was the traditional sunspot number data
(SN) from the World Data Center SILSO, Royal Observatory of Belgium,
Brussels\footnote{\url{http://www.sidc.be/silso/datafiles}}.
We note that these data have recently been subject to some corrections
that especially affect the low-frequency parts
{\citep{Clette2015,Clette2014,Usoskin2016,Lockwood2016}. Nevertheless, we use the older
calibration of the data set here, while we may return to the recalibrated
data set in a future publication.

Moreover, we deliberately left out the proxy set of the so-called group sunspot numbers \citep{Hoyt1998},
which dates back to 1610 and which is considered to be more reliable than the SN set. In addition to the so-called Dalton
minimum contained in the SN dataset, it also contains the so-called Maunder
minimum, the grandest solar minimum known from sunspot data. 
Because
this extreme minimum is included, the basic stationarity assumptions for
hindcasting cannot be considered valid.

For one particular demonstration we use some shorter proxy data
sets. They are too short to be useful in the hindcasting context. The RADIO
proxy are the 10.7 cm solar radio flux data \citep[][and references therein]{Tapping2013} downloaded from Laboratory for
Atmospheric and Space Physic (University of Colorado, Boulder) Time
Series Server
\footnote{\url{http://lasp.colorado.edu/lisird/tss/}}.

The MGII proxy is a composite
Mg II Index \citep[][and references therein]{Snow2014,Viereck2004} from the The Global
Ozone Monitoring Experiment webpage at the Institute of Environmental
Physics, University of Bremen\footnote{\url{http://www.iup.uni-bremen.de/gome/gomemgii.html}}.

The LYMAN proxy is a series of Lyman-alpha irradiance measurements \citep{Woods2000}
downloaded from the LASP Interactive Solar Irradiance Data Center
\footnote{\url{http://lasp.colorado.edu/lisird/lya/}}.

 We note that the PSI, SA, and SN all describe the sunspot
  (blanketing) component, while the MGII and LYMAN datasets are proxies
  of the facular (brightening) component. 
  In the physics-based models,
  the two most important proxy components are spots and faculae \citep[see, e.g.,][]{Yeo2014}. 
  We here concentrate our analysis
  on models that include only the spot component, and only use the MGII and LYMAN
to investigate the FDCs between the proxy pairs.

\subsection{Targets}\label{ss:targets}

Our target data sets, to be approximated by proxies, are all well known. For a detailed description of their differences we refer to 
\citet{Yeo2014b}.

The first data set we used as a modeling target is the ACRIM composite \citet{Willson2014}
\footnote{\href{http://www.acrim.com/Data Products.htm}{http://www.acrim.com/Data Products}}.

The second target data set is PMOD composite \citep[][for details]{FroehlichLean98}  and  \citep[][for updates]{Froehlich2006}\footnote{\url{ftp://ftp.pmodwrc.ch/pub/data/irradiance/composite/}}. 

The third data set we used as a target is an alternative compilation (RMIB, in some sources IRMB) from
the Royal Meteorological Institute of Belgium \citep{DeWitte2004,Mekaoui2008}\footnote{\url{ftp://gerb.oma.be/steven/RMIB_TSI_composite/}}.
}

The compiled values for all three targets are given in $Wm^{-2}$. 
However, their mean levels are different. We return to this aspect of the data sets below. 

We are quite aware that there is still a persisting controversy concerning
the differences especially in the decadal trends seen in the different
composites, as described, for instance, by \citet{Willson2014,Kopp2014}. Our goal here is to introduce a new data analysis method and report on the
new insights that this method can give to the ongoing discussion.
In the computational and graphical examples we most often use the
PMOD composite as a target and PSI as a proxy.

\section{Method}\label{s:method}
\subsection{Motivation}
Our simple method is a stepwise enrichment of a rather old and
simple idea: a combination of the input time series and their smoothed variants into one and the same regression model
(see, e.g., \citet{FroehlichLean98,Lean2000}). When Lean modeled the solar irradiance
in a semi-empirical way, he added
a term to the regression model. This component was ``smoothed over about 3 months''
  and described brightening
due to the faculae.
In this way, the regression scheme contained
the original component as well as its smoothed version. This approach
helped to improve the quality of the modeling. However, the exact method
of smoothing and the reason for using this particular amount of smoothing
was left open in the original paper. We aim to contribute to this point here. We introduce a particularly useful
smoothing scheme and determine proper parameters for
this smoothing.
\subsection{General scheme}
The typical prediction and hindcasting procedure consists of two
stages: model building, and application of the model. 
The model components are
available proxies or certain modifications of them. Below we use
the following notations. The target data (typically TSI composites in this
paper) are $y(t_i),i=1,\dots,N$. Proxy data sets are denoted as
$x(t_j),j=1,\dots,M$. The proxy data set spans a longer data interval
than the target data.  All the used models are in the form of linear
compositions $C(t_k)=\sum_{l=0}^L a_l E_l[\dots](t_k),k=1,\dots,K,$ where
$E_l[\dots](t_k)$ are the input values $x(t_k)$, transformed in certain ways. 

The exact parametrization (in square brackets), form, and nature of the
transformations are specified below.
The index $k$ runs over time moments common to both data sets, and the
coefficients $a_l$ are determined by minimizing the sum of squares $S$:
\begin{equation}
S=\sum_{i=1}^K \big((y(t_i)-C(t_i)\big)^2 = \sum_{i=1}^K \big(y(t_i)-\sum_{l=0}^L a_l E_l[...](t_i)  \big)^2.  
\label{LSQ} 
\end{equation}

It must be stressed here that all the model components are
first calibrated against target data sets by computing coefficients 
for the corresponding regression models. 
After calibration, the model can be evaluated for and compared to
the measured values of the target. In this case, we talk about
{\it \textup{prediction}}. When we evaluate the model for time points where only
proxy values are available, we perform {\it \textup{hindcasting}}.  
Some authors \citep[e.g.,][]{Herrera2015}
have postulated particular abstract components (e.g., wavelets or harmonics)
and used them after calibration to hindcast to past times, where
no proper data are available. We are significantly more
conservative here.

The quality of the models is evaluated using Pearson's correlation
coefficient $R_c$ , which is computed between the actual target data
and the predicted model values. 

For all correlations described below, we computed the $R_c$ value using only
time points where both arguments are available (excluding gaps in the two curves we studied).
We present our results with four-digit precision to reveal details of convergence of the computational iteration process and
details that are due to the high level of correlatedness between some computed solutions. Under current metrological
circumstances, this precision is
significantly higher than the actual measurement precision, and the results
can be interpreted accordingly.      

\subsection{Smoothing and detrending}

Our algorithms are based on smoothing and/or
detrending methods that are used in different contexts. From the wide
range of possible algorithms (moving average, weighted moving average,
least-squares spline approximation, etc.), we chose the classical Fourier
transform method. We 
transformed the input data (as is often
said, from the time domain to the Fourier domain), multiplied the
transformed data by a filter transfer function, and finally performed an
inverse transform. The result is a smoothed version of the original
input curve in time domain.
We used the Gaussian transfer function for smoothing:
\begin{equation}
T(\nu)=\exp^{-(\nu W)^2},
\end{equation}
where $\nu$ is the frequency in cycles per day, and $W,$ the width
parameter of the smoothing window, is 
here and throughout noted in units of day.
The width parameter $W$ characterizes the effective length of the
filter in the time domain, and correspondingly, $\delta\nu = 1/W$ is the
bandwidth in the Fourier domain. If the band is very narrow,
then the corresponding $W$ parameter is quite large, and straightforward filtering (convolving) in time domain becomes time
consuming. By using the Fourier transform method implemented as fast
Fourier transform, the computing time is significantly reduced.

\begin{figure} 
\centerline{\includegraphics[width=0.5\textwidth,clip=]{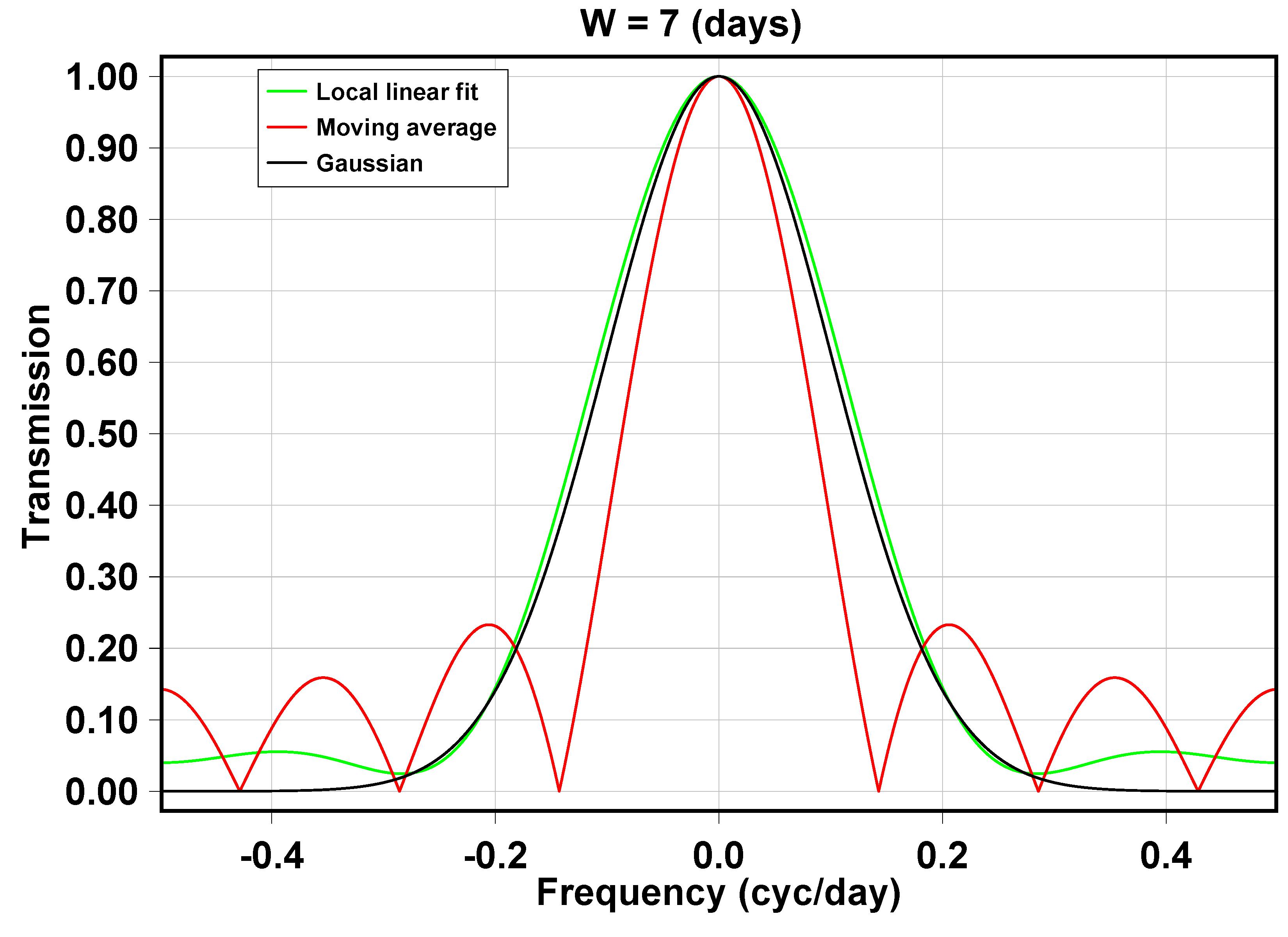}}
\caption{Three different transmission curves for a specific width of the time
  domain filter.}\label{fig001}
\end{figure}

Our
method is equivalent to smoothing the original data with a
Gaussian window in time domain \citep[used, e.g., in][]{Ball2014}. The particular method of parametrization is chosen
so that the corresponding smoothing effect can be easily compared with
traditional moving averaging.
The Fourier smoothing width $W$ is therefore comparable to the $W$-day moving
average.
In Fig.~\ref{fig001} we show the transmission functions of three different smoothing
filters: a local linear fit, a running average, and a Gaussian
filter with the width $W=7$ .  The
frequently used moving-average filter clearly has several parasitic sidebands,
the local linear fit has one sideband, and 
the Gaussian filter lacks any sidebands.
From this it follows that the moving average is not the best device to cut off
high-frequency noise or separate different frequency bands in time
series. The downweighted local linear fit method (which is also a widely used smoothing method) approximate transmission curve is quite
similar to the Gaussian having only minor extra transmissions in the
region of the higher frequencies.

The Gaussian-smoothed
versions of the particular input proxies are denoted by $E[W,0]$, where $W$ is filter width
parameter, and the zero as the second parameter stands for no offset
applied; this parameter is non-zero for the passband filters introduced
in Sect.~\ref{ss:bandpass}). For generality, we also use the notion
$E[0,0](t)=x(t)$ for the original signal.
The detrended version of the input proxy
$x(t_j)-E[W,0](t_j),j=1,\dots,M$ is denoted by
$E_d[W,0]$. Detrending allows us to emphasize the features in the high-frequency regions of the data.

\subsection{Bandpass filtering}\label{ss:bandpass}
The smoothing process itself is a low-pass filtering in terms of
signal processing theory. In the context of the Fourier transform
method, we can also consider the so-called bandbass filters. The
transformed signal is multiplied by a transfer function that consists
of two symmetrically placed passbands.
We used the
simplest bandpass filters with the transfer function
\begin{equation}
T(\nu)=\exp^{-((\Delta \nu -\nu) W)^2}+\exp^{-((\Delta\nu +\nu) W)^2}, 
\end{equation}  
where $\Delta\nu = 1/O$ is the passband offset. Below we use two
parameters for bandpass filtering, the width parameter $W,$ and the offset
parameter $O$, both measured in days. The $O$ parameter is
essentially the {\it \textup{period}} of the waveforms that pass through the
filter unchanged (it determines the positions of the two maxima of
the transfer function).
For the bandpass-filtered time-dependent regression components we added the
$O$ parameter to the general expressions - $E[W,O](t)$.
\begin{figure} 
\centerline{\includegraphics[width=0.5\textwidth,clip=]{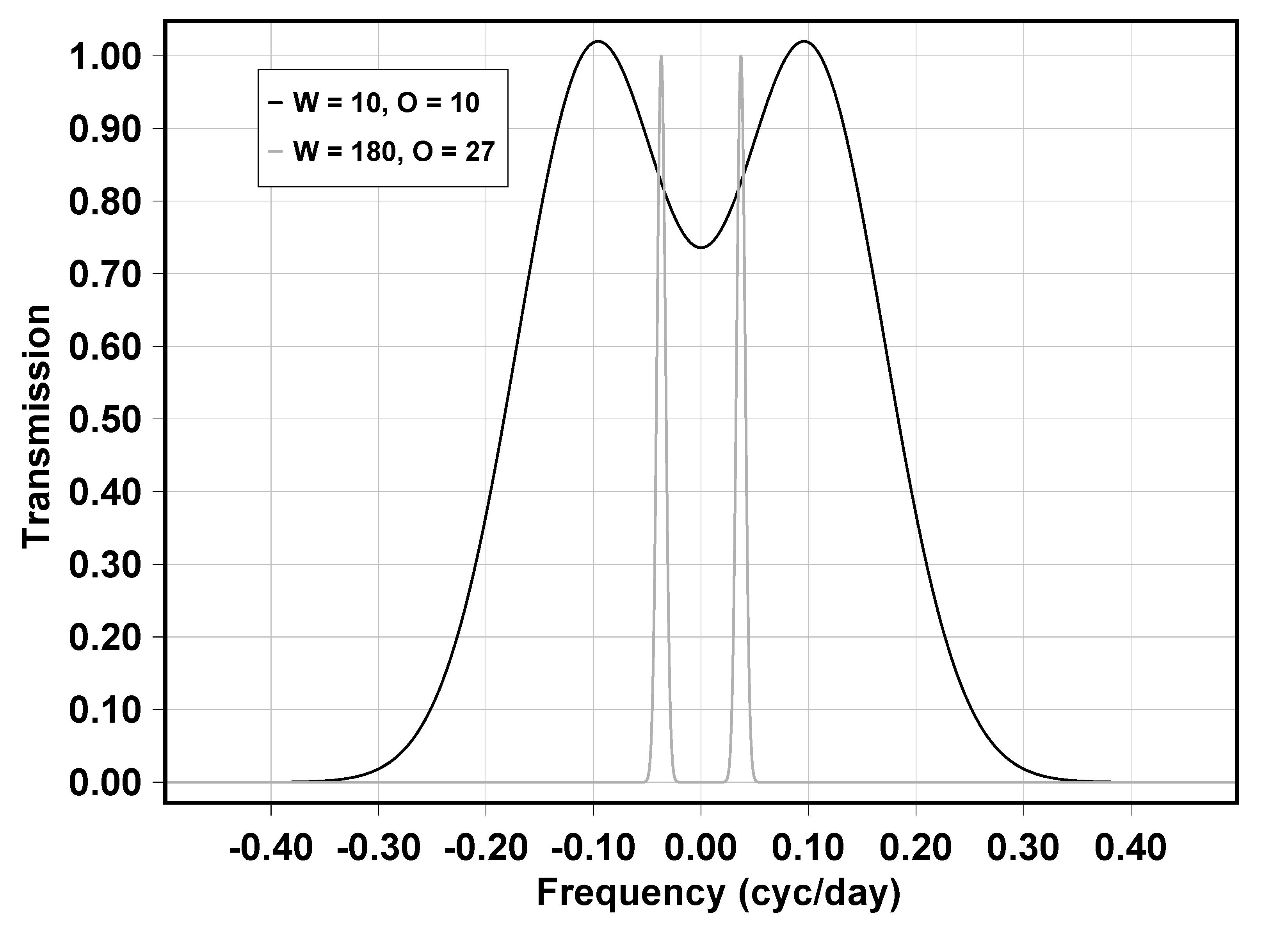}}
\caption{Two different transmission curves for bandpass filters.}
\label{fig002}
\end{figure} 
In Fig.~\ref{fig002} we plot the two transmission curves for the bandpass
filters. We used the differently filtered time series
as regression predictors, which means that the exact
normalization of the transfer functions is not important. The
amplitude differences are absorbed into regression coefficients.

\subsection{Envelopes}\label{ss:envelopes}

When we smooth the input data with different smoothing window parameters
$W_i,O_i,i=1,\dots$ we obtain a set of curves that we can use as
components for the regression modeling.  For instance, if we
systematically build a set of smoothed curves with offsets $O_j=1/ (j \cdot \delta
\nu_0),j=1,\dots,J$ where $\delta \nu_0$ is the frequency step, then by
choosing proper $\delta \nu_0,$ we can well approximate the method of
convolution kernel fitting, see, for instance, \citet{Preminger2005}. The
important point here is that regression on smoothed components or
approximation by moving kernel (convolution) are both fully linear
procedures, that is, the predictions depend linearly on 
input data or
smoothing kernel values.

To widen the range of modeling possibilities, we introduced a mild
nonlinearity into our components. This allowed us to take
the ``sidedness'' of the involved correlations into account and to move information along
the frequency axis.
This is useful, for instance, when
the faster changing blanketing effect of the sunspots  should influence the much slower changes in the overall network brightness. 

\begin{figure} 
\centerline{\includegraphics[width=0.5\textwidth,clip=]{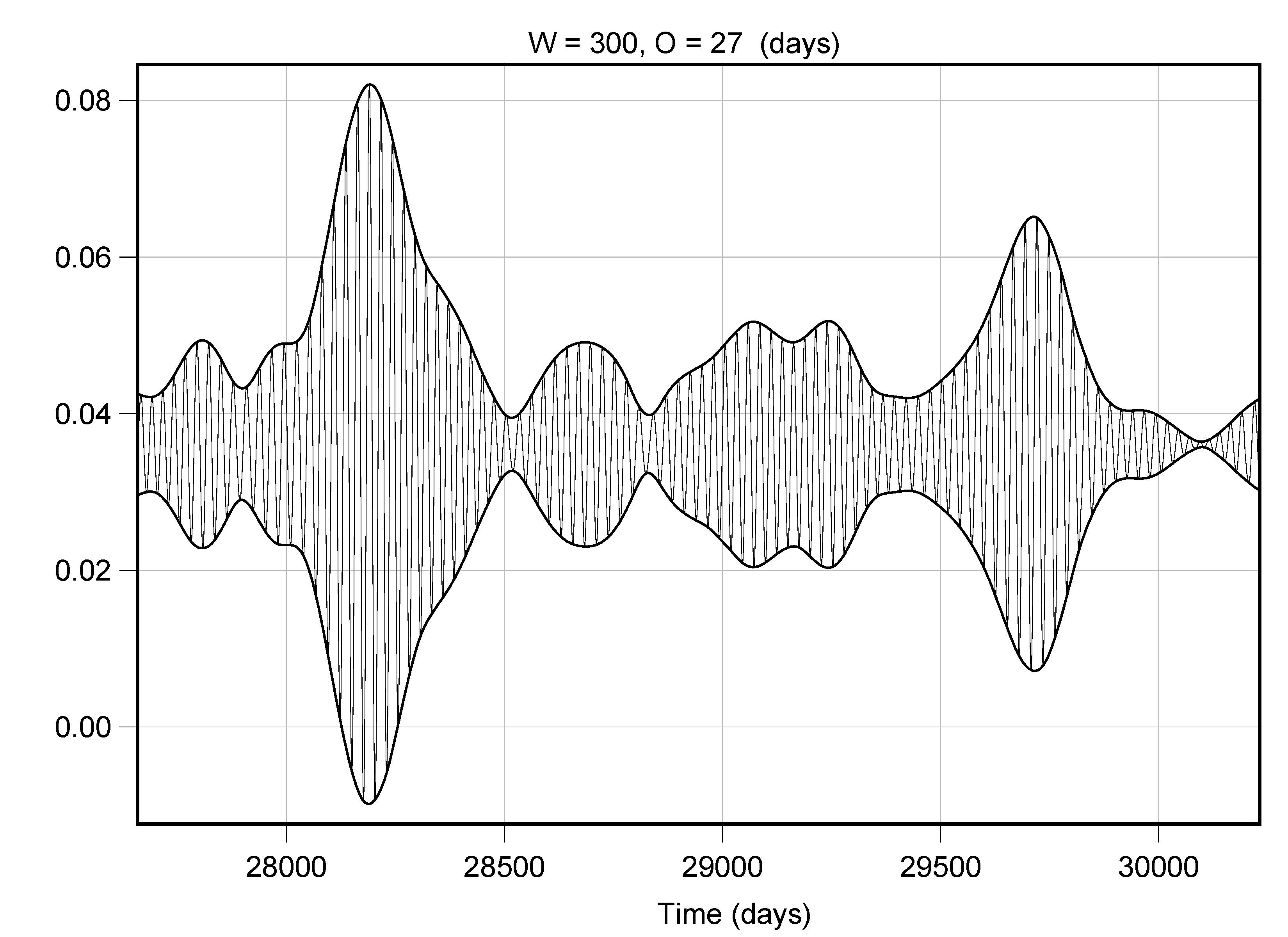}}
\caption{Fragment of bandpass filtered PSI data set $E[W,O](t)$ with upper $E^{+}[W,O](t)$ and lower $E^{-}[W,O][t]$ envelopes. The envelopes are much smoother and have a sidedness that can take correlations or anticorrelations into account.}
\label{fig003}
\end{figure}

One of the simplest nonlinear transforms of this type is the computation of
envelopes.
In this approach, we filter the input data set with a bandpass filter (e.g., with parameters $W$=300d and $O$=27d) and then compute upper and lower
envelopes for the filtered data (see Fig.~\ref{fig003}).
The envelopes take the sidedness of different effects into account.
They are also significantly smoother (shifting information in the
Fourier domain from high to low frequencies).

There are many methods to estimate smooth envelopes but one of the simplest
one is a spline interpolation through maxima (or minima). This is the method used in this paper.
As seen from
Figure~\ref{fig003} the sets of the extrema are well defined for
the band-pass filtered signals.

Our method is somewhat similar to the method of empirical mode
decomposition that has been used in the same context \citep[see, e.g.,][]{Barnhart2011,Li2012}. 
For the various components we obtained by bandpass filtering,
we can find rather similar intrinsic mode functions. However, we
prefer the somewhat simpler Fourier analysis approach because here the
spectra of the important modes are highly concentrated in the frequency
domain. In the solar context at least part of the variability is
coherently clocked by the rotation. Quite close to our approach,
at least from the methodological point of view, is the use of so-called wavelets and cross-wavelets \citep[see, e.g.,][]{Benevolenskaya2014,Xiang2014}. 

\citet{Rypdal2012} used amplitude detrending together with mean deterending to reveal stationary (or
statistically stable) fluctuations.  Envelopes, as we
introduced above, can be used with the same goal.
\begin{figure} 
\centerline{\includegraphics[width=0.5\textwidth,clip=]{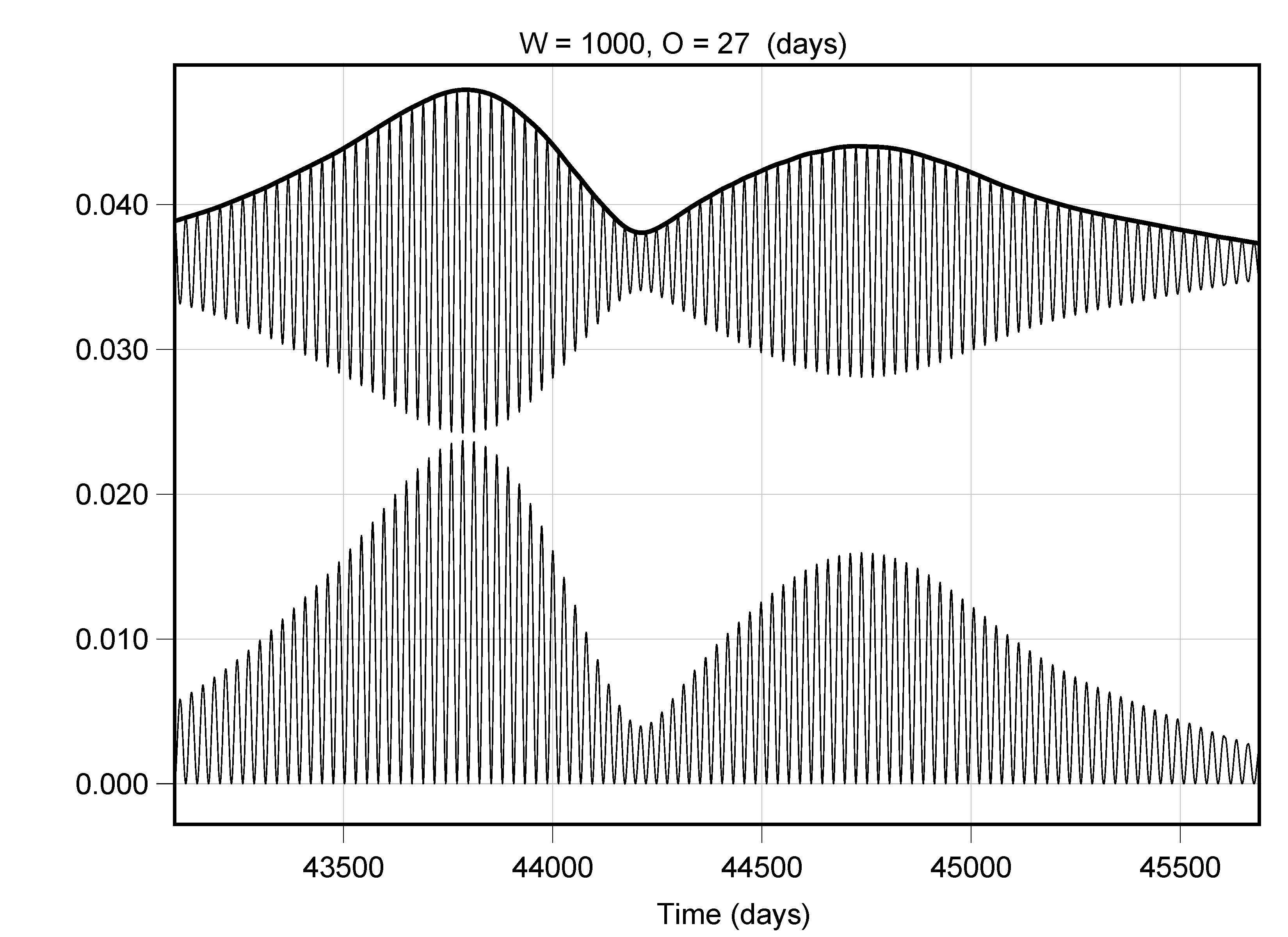}}
\caption{Fragment of bandpass-filtered PSI data set} $E[W,O](t)$ (thin curve above), its upper
envelope $E^{+}[W,O](t)$ (thick curve), and the corresponding detrended version
$E^{+}_{d}[W,O](t)=E[W,O](t)-E^{+}[W,O](t)$ (thin curve below). 
\label{fig004}
\end{figure} 

For the different smoothed (filtered) time-dependent data sets
we used the following simple notions: $E[W,O](t)$ for bandpass
filtered data, $E^{+}[W,O](t)$ for the upper and $E^{-}[W,O](t)$
for the lower envelopes.
The additional index $d$ for
$E$ means that instead of the smooth envelopes of the curve being involved,
they are used to detrend the bandpass-filtered data, or formally
$E_{d}^{+}[W,O](t)=E[W,O](t)-E^{+}[W,O](t)$ for detrending with
an upper envelope (see Fig.~\ref{fig004}).
\begin{table*}[ht!]
\caption{Correlation matrices for input proxy data sets.}
\begin{center}
\begin{tabular}{lrrrrrr}
  & PSI & SA & SN  & RADIO & MGII & LYMAN \\
\hline
Original & & &  \\
\hline
PSI &     1.000 &     0.942 &     0.852 &     0.872 & \iti{     0.764 }  &     0.771 \\ 
SA &     0.942 &     1.000 &     0.879 &     0.901 &     0.792 &     0.812 \\ 
SN &     0.852 &     0.879 &     1.000 &     0.946 &     0.915 &     0.912 \\ 
RADIO  &     0.872 &     0.901 &     0.946 &     1.000 &     0.950 &     0.954 \\ 
MGII  & \iti{     0.764 }  &     0.792 &     0.915 &     0.950 &     1.000 & {\bf     0.972 }  \\ 
LYMAN  &     0.771 &     0.812 &     0.912 &     0.954 & {\bf     0.972 }  &     1.000 \\ 
\hline
LF & & &  \\
\hline
PSI &     1.000 & {\bf     0.998 }  &     0.989 &     0.994 &     0.989 &     0.990 \\ 
SA & {\bf     0.998 }  &     1.000 &     0.988 &     0.988 & \iti{     0.982 }  &     0.982 \\ 
SN &     0.989 &     0.988 &     1.000 &     0.991 &     0.986 &     0.984 \\ 
RADIO &     0.994 &     0.988 &     0.991 &     1.000 &     0.994 &     0.992 \\ 
MGII &     0.989 & \iti{     0.982 }  &     0.986 &     0.994 &     1.000 &     0.990 \\ 
LYMAN &     0.990 &     0.982 &     0.984 &     0.992 &     0.990 &     1.000 \\ 
\hline
HF & & &  \\ 
\hline
PSI &     1.000 & {\bf     0.894 }  &     0.741 &     0.814 &     0.542 &     0.586 \\ 
SA & {\bf     0.894 }  &     1.000 &     0.745 &     0.821 & \iti{     0.521 }  &     0.591 \\ 
SN &     0.741 &     0.745 &     1.000 &     0.795 &     0.657 &     0.666 \\ 
RADIO &     0.814 &     0.821 &     0.795 &     1.000 &     0.741 &     0.791 \\ 
MGII &     0.542 & \iti{     0.521 }  &     0.657 &     0.741 &     1.000 &     0.862 \\ 
LYMAN &     0.586 &     0.591 &     0.666 &     0.791 &     0.862 &     1.000 \\ 
\hline
\end{tabular}
\tablefoot{The Original column presents correlations between the input proxy data, LF presents correlations after applying
a Gaussian smoothing filter with the width parameter $W=750,$ and HF consists of correlations between detrended proxies. Here and below the maximum values 
for each subtable are given
in boldface and minimum values in italics.}
\label{table002}
\end{center}
\end{table*}

After defining our rather simple tool set of low-pass filtering
(smoothing), band-pass filtering,
envelope building, and detrending, we  demonstrate their usefulness
below in various data processing contexts.

\section{Diagnostic tests}\label{s:diagnostics}

Using the simplest smoothing and detrending operators, we can obtain some insight into 
the physical characteristics of and problems related to different input proxy and target data sets. 
First we divide the data sets into two different parts, a smoother LF part, and a
faster changing HF part. By systematically correlating these smoothed or detrended parts, we can better characterize the potential problems with data. Then we
use parameter-dependent bandpass smoothing to build spectra that help to localize
correlating and anticorrelating frequency bands of the data sets.  
These two simple exploratory type diagnostic tests allowed us to set up the general scene 
for the further investigations below.   
\subsection{FDCs of proxies and targets}
The proxy and target time series all have prominently visible changes
over distinct frequency ranges: a short-term variation, changes over
the scale of the solar cycle, and possibly also some secular
trends. To start, we therefore used simple Gaussian smoothing with the width
parameter, $W_0$, for all data sets to separate the
signals into an LF and an HF component. The LF component is
an input proxy data set smoothed with the Gaussian filter $E[W_0,0](t),$
and the HF component is its detrended variant
$E_d[W_0,0](t)=E[0,0](t)-E[W_0,0](t)$.
The limiting width $W_0=750$ was chosen using an optimization procedure of
the first (and most prominent) components of the different regression
models (see Sect.~\ref{s:simple} for a detailed analysis).
The results do not depend very much on this
particular choice (we tried values between 500-1000 days). For
simplicity, we call the $W$ parameter in this simple analysis
scheme a ``breakpoint'' and the LF part of the data sets a
``backbone''.

First we examine the correlation matrices between the various proxy
data, correlating the original proxies without smoothing, and the
LF/HF parts separately; the results are presented in Table~\ref{table002}.
The LF components of the proxies are very highly
correlated ($R_c = 0.982-0.998$), even those that describe 
very different features (e.g., PSI vs. MGII). In other words, they can
confidently be treated as interchangeable. This is not so for the HF part
correlations: while PSI and SA correlate strongly with $R_c=0.894$
(because they are essentially
only slightly different measures of the spotedness), the correlation
between the SA and MGII index, for example, is
significantly lower ($R_c=0.521$).
In physical terms we can see that all the proxies describe essentially
the same smooth (LF) change in the activity level, but the HF fluctuations
are different, especially when the two different types (blanketing
vs. brightening) of proxies are compared.
This simple insight is used below, where we build different target approximations from proxy-based components. 

\begin{table}[ht!]
\caption{Correlation matrices for input target sets.}
\begin{center}
\begin{tabular}{lrrr}
  & PMOD & RMIB & ACRIM  \\
\hline
Original & & &  \\
\hline
 PMOD &     1.000 & {\bf     0.937 }  &     \iti{ 0.861}  \\ 
 RMIB & {\bf     0.937 }  &     1.000 &     0.906  \\ 
 ACRIM &     \iti{ 0.861} &     0.906 &     1.000   \\ 
\hline
LF & & &  \\
\hline
PMOD &     1.000 &     {\bf 0.917} &     \iti{ 0.849}  \\ 
RMIB &     {\bf 0.917} &     1.000 &     0.897  \\ 
ACRIM &     \iti{ 0.849} &     0.897 &     1.000  \\ 
\hline
HF & & &  \\
\hline
PMOD &     1.000 & {\bf     0.957 }  &     \iti{ 0.906}  \\ 
RMIB & {\bf     0.957 }  &     1.000 &     0.921  \\ 
ACRIM &     \iti{ 0.906} &     0.921 &     1.000   \\ 
\hline
\end{tabular}
\tablefoot{
The same as for Table~\ref{table002}.    
}
\label{table003}
\end{center}
\end{table}

\begin{figure} 
\centerline{\includegraphics[width=0.5\textwidth,clip=]{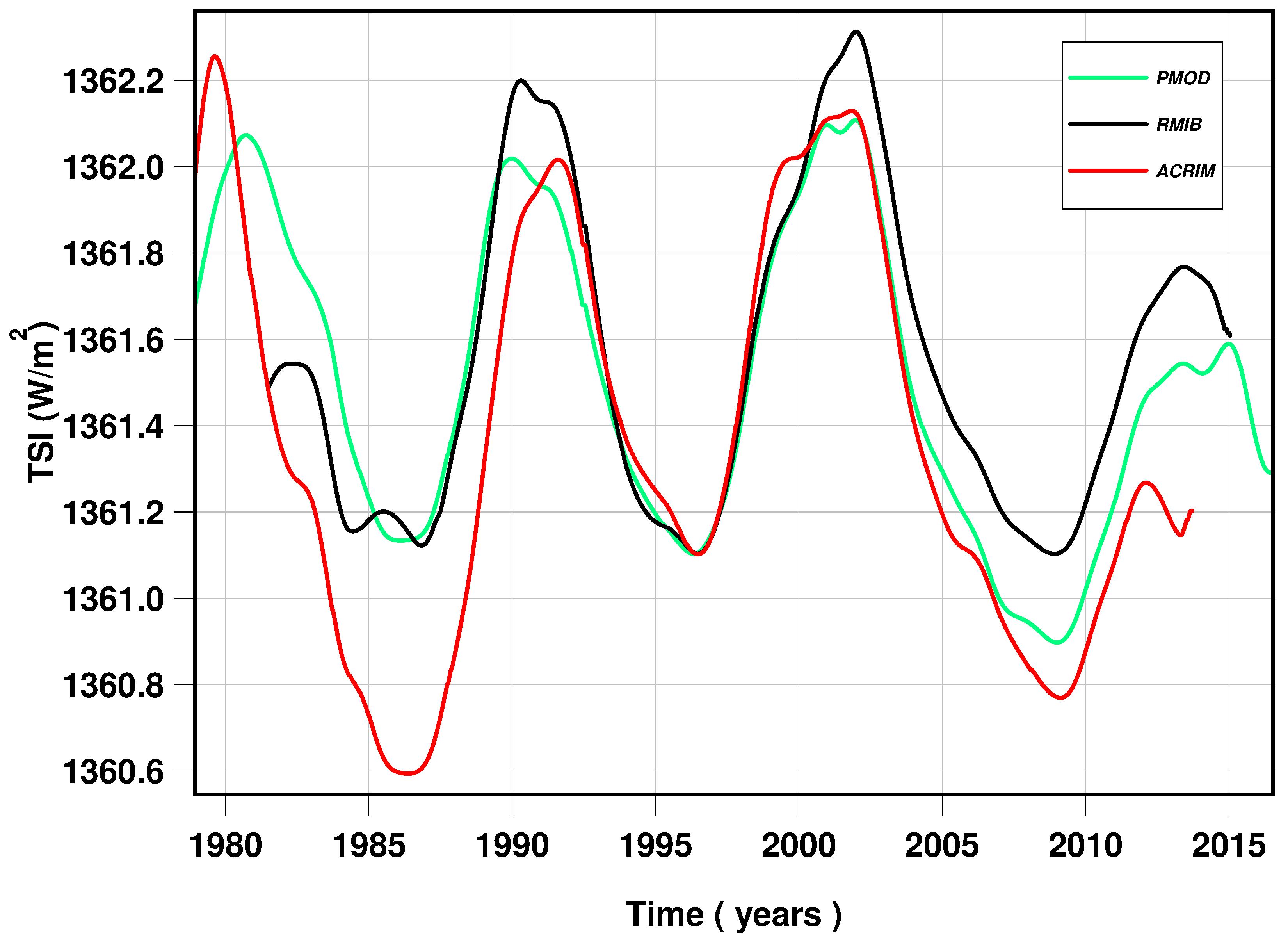}}
\caption{Main controversy. To recover TSI for the past dates, we need reliable current estimates for calibration.
The LF parts (obtained with a Gaussian smoothing filter with $W_0=750$) of
three 
target TSI composites differ
significantly, however. The 
curves are shifted to a common level at
1996.465.}
\label{fig005}
\end{figure}

Next, we investigate the correlations between the target curves with
the same technique, and present our results in Table.~\ref{table003}.
As evident from this table, the structure of the correlations between
the targets is different,
the almost perfect correlation of the LF backbone seen in the proxies
has significantly decreased for the targets.
The range of variability is rather similar for all three comparison sets. For the original target data sets the
correlation range is $0.861-0.937$, for the LF parts it is
$0.849-0.917,$ and for the HF components it is $0.906-0.957$.
 For the PMOD-RMIB pair the HF correlation is somewhat higher than
  for PMOD/RMIB-ACRIM, which is also expected because PMOD and RMIB are
  based on the same measurements.
  The stronger differences in the LF backbones in Fig.~\ref{fig005} reflect issues in the
  TSI composite building, caused for example by the different methods used to take the instrument degradation trends or improper stitching of the
  observed fragments into account (see \citet{Kopp2014} for the assessment of problems involved). This
  main controversy is an expected result, 
  but we here express it explicitly and demonstrate
  the effects it has for model building.

The difference of the LF/HF behavior for the proxy and target data sets
is crucial. It shows that regardless of the linear combination methods we use,
the set of simple proxies is not sufficient to properly describe all
the target sets. The rather large variability in the LF components of the
target sets needs additional arguments to be explained and potentially
requires additional data to determine which of them should be used
for proper TSI reconstruction and hindcasting.

\subsection{Narrowband FDCs}

\begin{figure} 
\centerline{\includegraphics[width=0.5\textwidth,clip=]{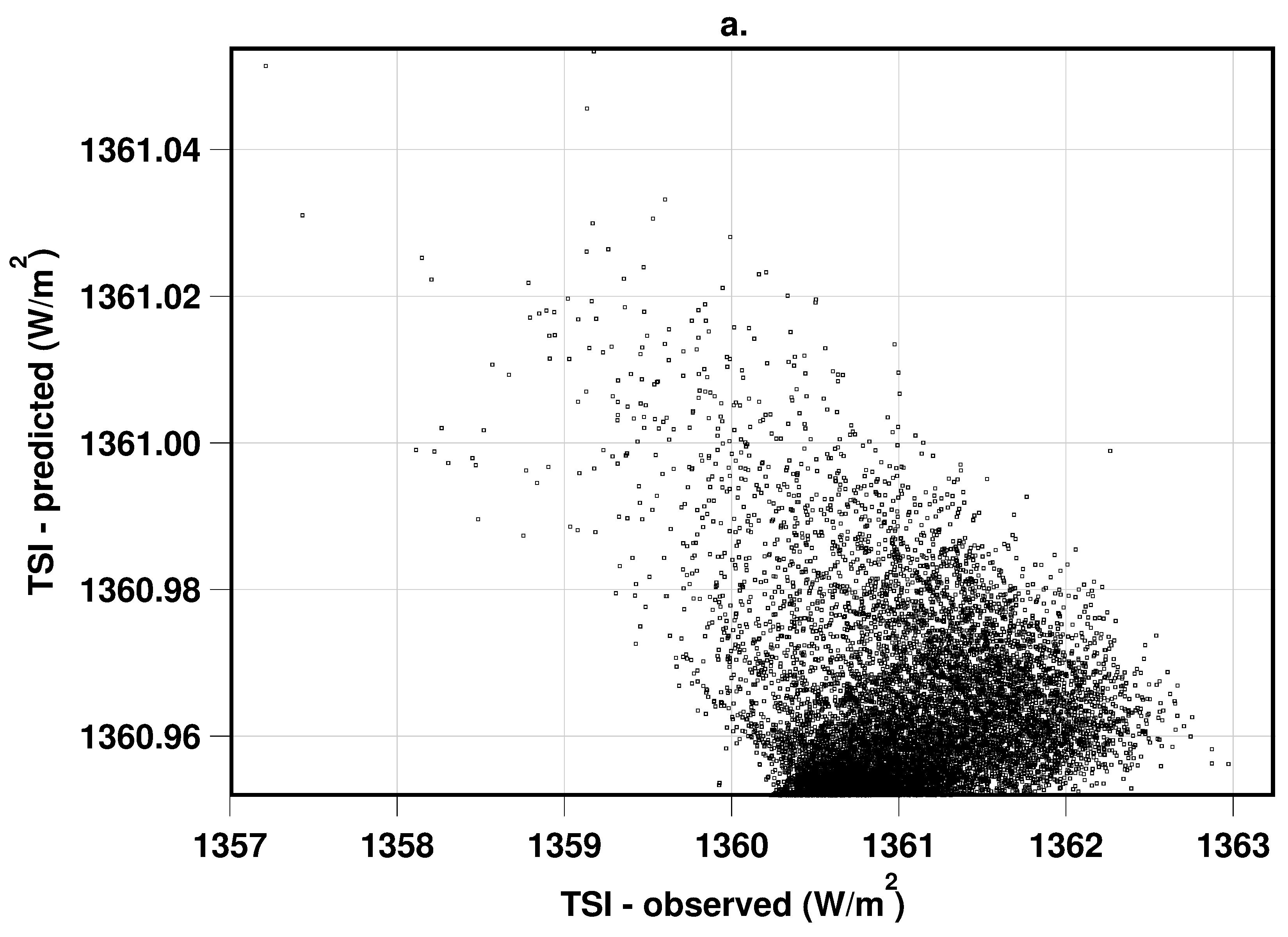}}
\centerline{\includegraphics[width=0.5\textwidth,clip=]{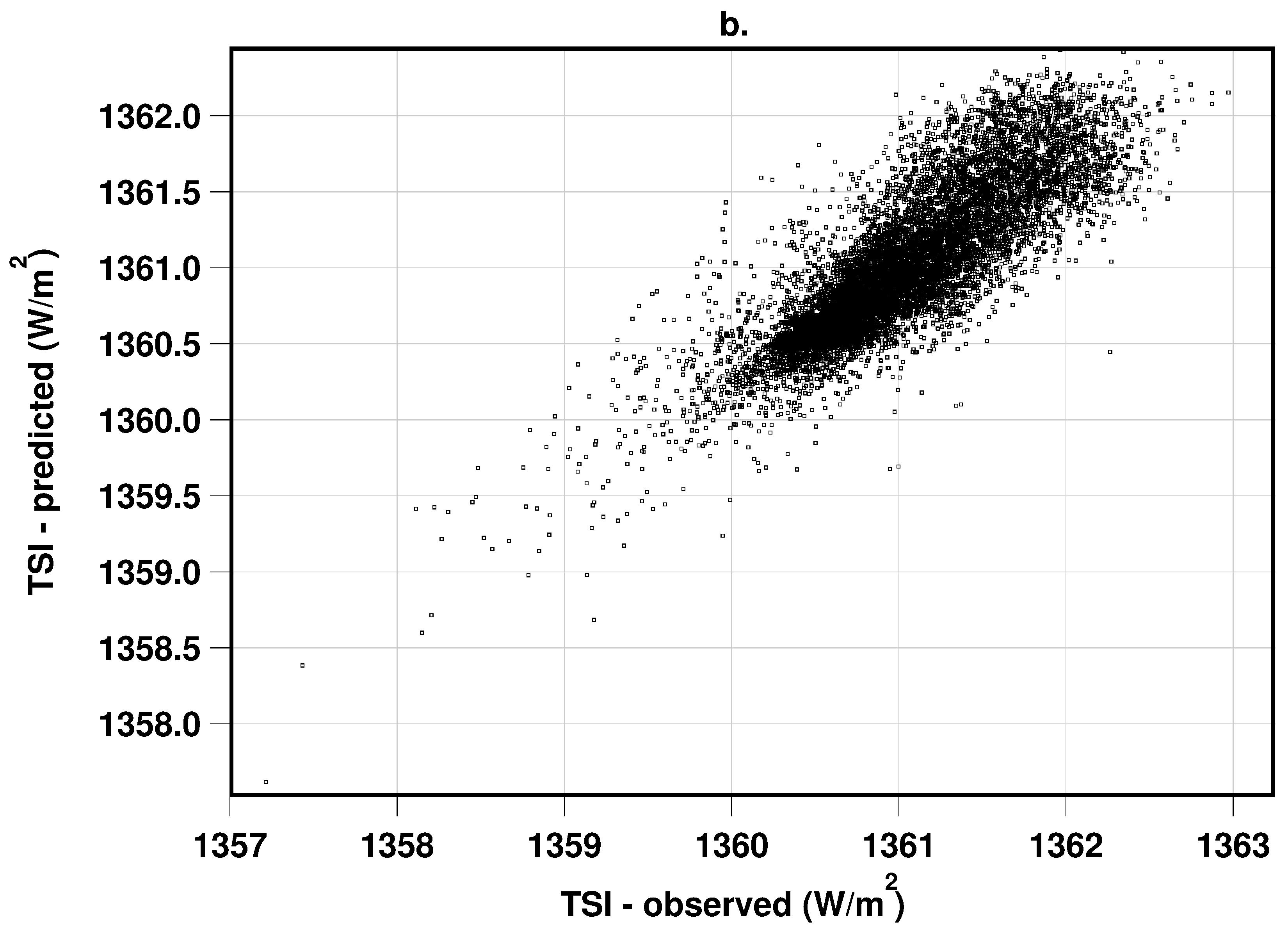}}
\centerline{\includegraphics[width=0.5\textwidth,clip=]{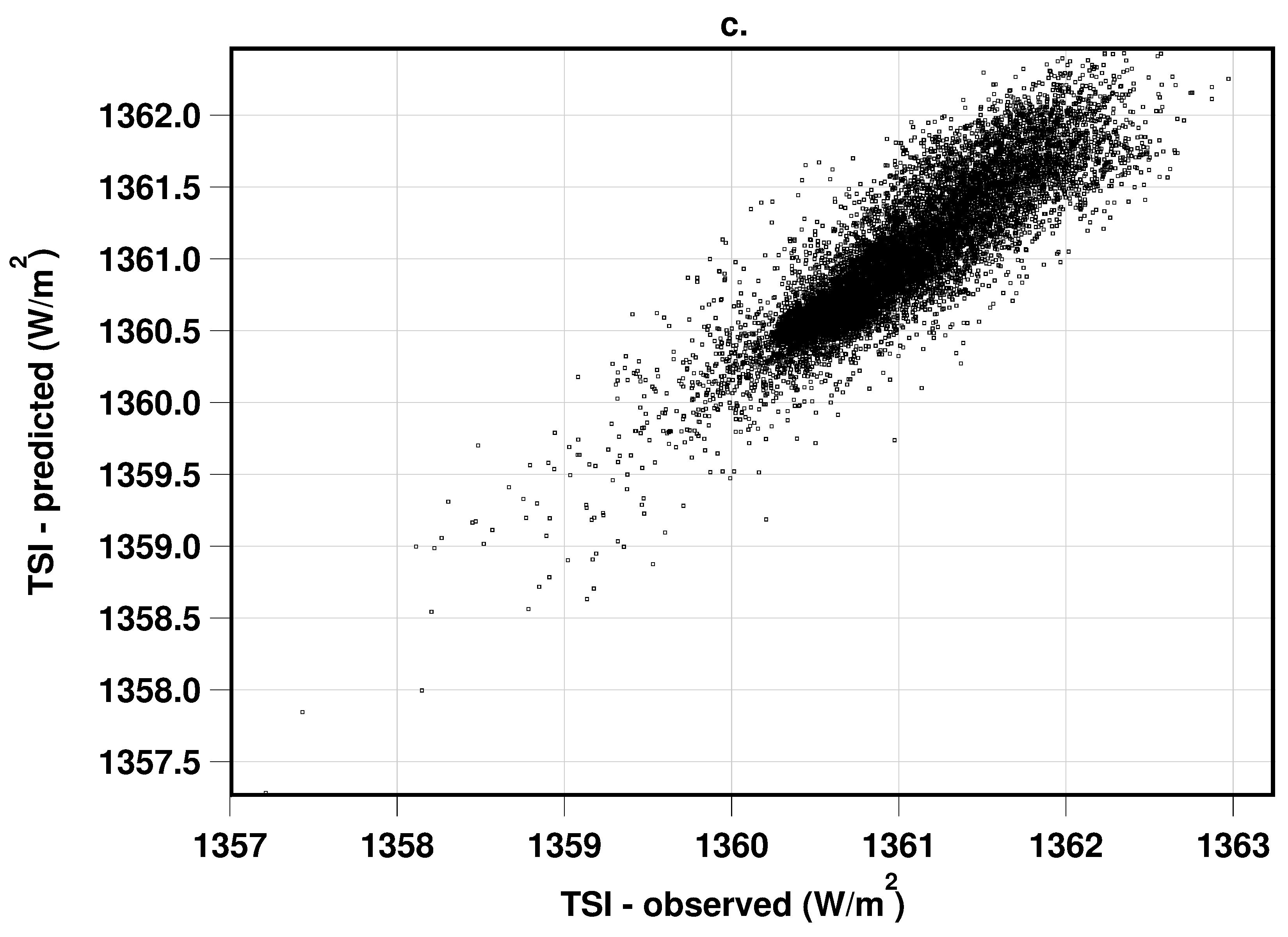}}
\caption{Scatter plots for PSI-PMOD pair: a. - linear fit of the proxy to data ($R_c = 0.020$); 
b. - regression modeling using the simple model ($R_c = 0.860$); c. - modeling using multicomponent model ($R_c = 0.893$).  
}
\label{fig006}
\end{figure}

It is well known that traditional activity indicators (e.g., sunspot
numbers) are not well correlated linearly with the TSI measurement series
\citep[see, e.g., a recent demonstration by][]{Hempelmann2012}. In Fig.~\ref{fig006}a we present
a typical scatter diagram, in this case between the overlapping parts of the PSI
and PMOD data.  We can try to build nonlinear regression curves
between these two, but the predictive power of this type of relations
is very low \citep{Preminger2005,Zhao2012,Hempelmann2012}.

\begin{figure} 
\centerline{\includegraphics[width=0.5\textwidth,clip=]{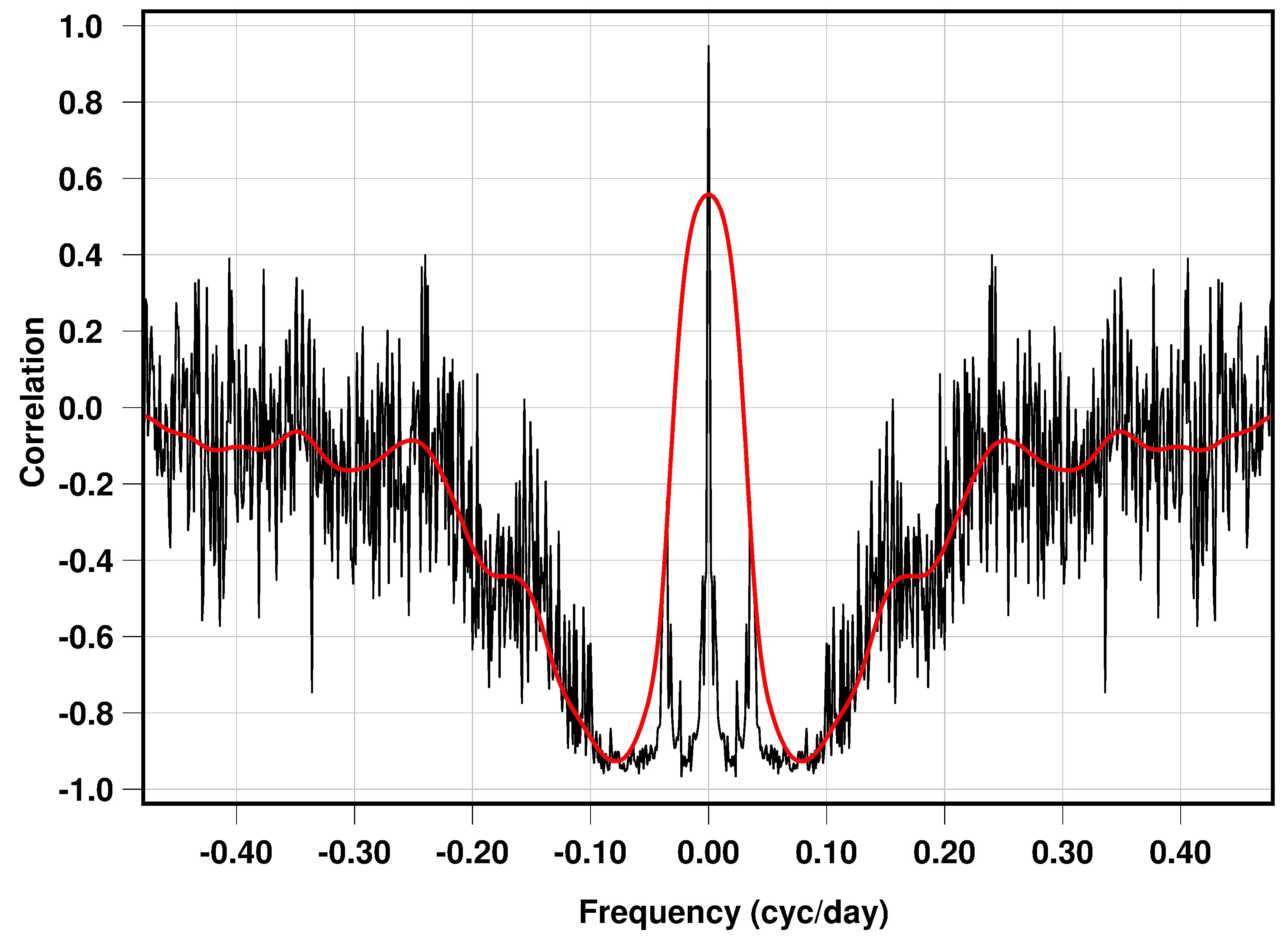}}
\caption{FDCs between overlapping parts of the PMOD and PSI data sets as function of the band-pass offset $\Delta\nu$.
Width} parameter $W=2000$ - black, $W=30$ - red.
\label{fig007}
\end{figure}

With a proper filtering and data analysis technique, we can
characterize our input data sets (e.g., PSI as proxy and PMOD as target)
in terms of the FDCs. This means we do not correlate original data
sets, but various parts of them in the frequency domain.  For this
purpose, we filter proxy and target curves with bandpass filters with
different frequency offset parameters and correlate the results (by
computing standard Pearson correlation coefficients $R_c$). 
In Fig.~\ref{fig007} we show the results of such a simple computation.
The filter width parameter $W=2000$ was chosen so that enough
detail would be revealed. This value is a good compromise between
frequency resolution and
lower signal-to-noise ratio
that is due to the narrowness of the filter. The spectrum for $W=30$ is discussed below.
Figure~\ref{fig007} clearly shows that different parts of
the spectrum correlate in different directions. First, there is
a strong central peak with a high positive correlation  that arises from the highly correlated LF components. Then
there is a wide band up to the frequencies approximately $0.1$ cycles
per day with a strong anticorrelation. From $0.1$ to $0.2$ a
transient band is visible where the correlations are still negative,
but no longer strong. Finally, our ``spectrum'' starts to wildly oscillate around the zero level correlation.  The
region of negative correlations also shows separate ``spectral
lines'' with a weaker anticorrelation. The most prominent
of these lines occurs near the solar rotational frequency $\nu = 1 /
27$d$^{-1}$.
This shows that the method of FDCs is rather sensitive to various
aspects of the TSI variability (in this case, to solar rotation).

The quite simple experiments with smoothing and bandpass
filtering allow us to reveal interesting aspects of the proxy - target
relations. A more systematic approach demands introduction of certain
regression schemes.

Motivated by the results of the diagnostic tests above, we build below different regression models
of the type of Eq.~(\ref{LSQ}). 

\section{Simplest regression models}\label{s:simple}

Because our approach contains a number of new
techniques and notions, it is useful to describe their application in
a sequence of steps, starting from the simplest steps, and finally showing
the final results.
 We start from the model where only one smoothed component, the
original proxy, and a
  constant level are used as regression components. Moreover, hereafter we only consider
  proxies related to the blanketing of sunspots (SA, SN, and PSI), and therefore we do not
  directly model the facular brightening component (that could be described, e.g., by MGII and LYMAN
  data sets). One important argument for this neglection arises from the shortness
  of the facular proxies, due to which they have a limited hindcasting
  capacity. 
  During our step-by-step approach, even without
  the facular component, our methods yield a modeling power 
  comparable to the physics-based models that include the brightening component.

This simple model can be useful as a poor man's modeling device.

\subsection{Three-component model}
The FDCs plot in Fig.~\ref{fig007} shows two main features: 
a highly correlated LF part (peak at the center), and
a wide anticorrelated band at higher frequencies. Based on these
characteristics, we assume that the following regression model can be
used to approximate the TSI using a proxy:
\begin{equation}
C(t)=a_0+a_1 E[0,0](t)+a_2 E[W,0](t),
\label{simplemod}
\end{equation}
where we use the general notions $E[0,0](t)$ for the original proxy and $E[W,0](t)$
for its smoothed variant. The smoothed component should correlate with
low
frequencies of the target curve, and the second term, with the coefficient
$a_1, a_1<0$, should account for the anticorrelations. The parameter
values of the model $C(t)$ were obtained by combining linear least-squares minimization for coefficients $a_0$, $a_1$ and $a_2$ with
an exhaustive grid search for parameter $W$.  The results for a concrete
example of the proxy/target pair, PSI vs. PMOD, are listed in
Table~\ref{table004}. 
\begin{table}
\caption{Regression model in standard format for the PMOD target built using PSI as a proxy.}
\begin{center}
\begin{tabular}{llrrc}
Coef. & Type & $W$ & $1/O$ & Value \\
\hline
$a_{0}$ & 1.0 & - & - &    1361 \\
$a_{1}$ & $E$ & 0 & 0 &      -8.004 \\
$a_{2}$ & $E^{ }$ &     791.9 &  -  &      19.08 \\
\hline
\end{tabular}
\label{table004}
\end{center}
\end{table}
The prominent feature of the simplest solution
is the minus sign for the coefficient $a_1$, which means that the original
non-smoothed component enters into the model in a reversed way, while the
smoothed component enters with a positive coefficient $a_2$. In the
context of our study, this is quite understandable because the spots serve
as radiation-blocking elements on the solar disk. The overall level of
solar magnetic activity is modeled by the smoothed with $W=792$
component.
Our simple model convincingly demonstrates how the correlating
and anticorrelating elements of the input curve can be separated by
using Gaussian smoothing with a properly set
smoothing window width.

In Fig.~\ref{fig006}b we show a scatter plot of the observed TSI (PMOD
composite) and our simplest PSI-based regression model, showing a prediction
modeling precision at correlation level $R_c=0.860$. As evident when comparing
with the correlation matrices for the targets in Table~\ref{table003},
the obtained value is practically the same as some correlations between targets  (e.g., ACRIM vs. PMOD at $R_c=0.861$).
Consequently, at the current level of observational precision, the errors
brought in by the modeling procedure are not significantly higher than
the variations between the input data sets. 

This urges us to proceed to other computational experiments.

\subsection{Baron von Munchausen method (BvM)}

Because our knowledge about the directly measured targets is controversial
in the sense that their LF parts differ significantly, we assume
that we can precisely estimate the real target TSI values with the
following data manipulation method, which we call Baron von
Munchausen method (hereafter BvM):
we take the TSI curve estimated from regression modeling (e.g., of the PMOD), subtract its
LF component, and add the LF component of the actual
target. This means that we try to ignore the errors that are
due to the improper
modeling of the LF part of the TSI. In the particular case of modeling PMOD
using PSI as a proxy, we obtain the following result:
while the resulting model correlation for the real data is $R_c=0.861$,
for the combined curve in which the LF part of the
regression model is replaced with the LF part of PMOD (BvM), it
increases to $R_c=0.887$,
see Table~\ref{table006}. This is the potential prediction accuracy level
for the simplest model for the particular case when the following two conditions are fulfilled:
the LF part of the target TSI is measured correctly, and the LF part of the
target TSI does not contain any secular components (all its variability is strongly connected to
the magnetic tracer statistics). 

\begin{figure} 
\centerline{\includegraphics[width=0.5\textwidth,clip=]{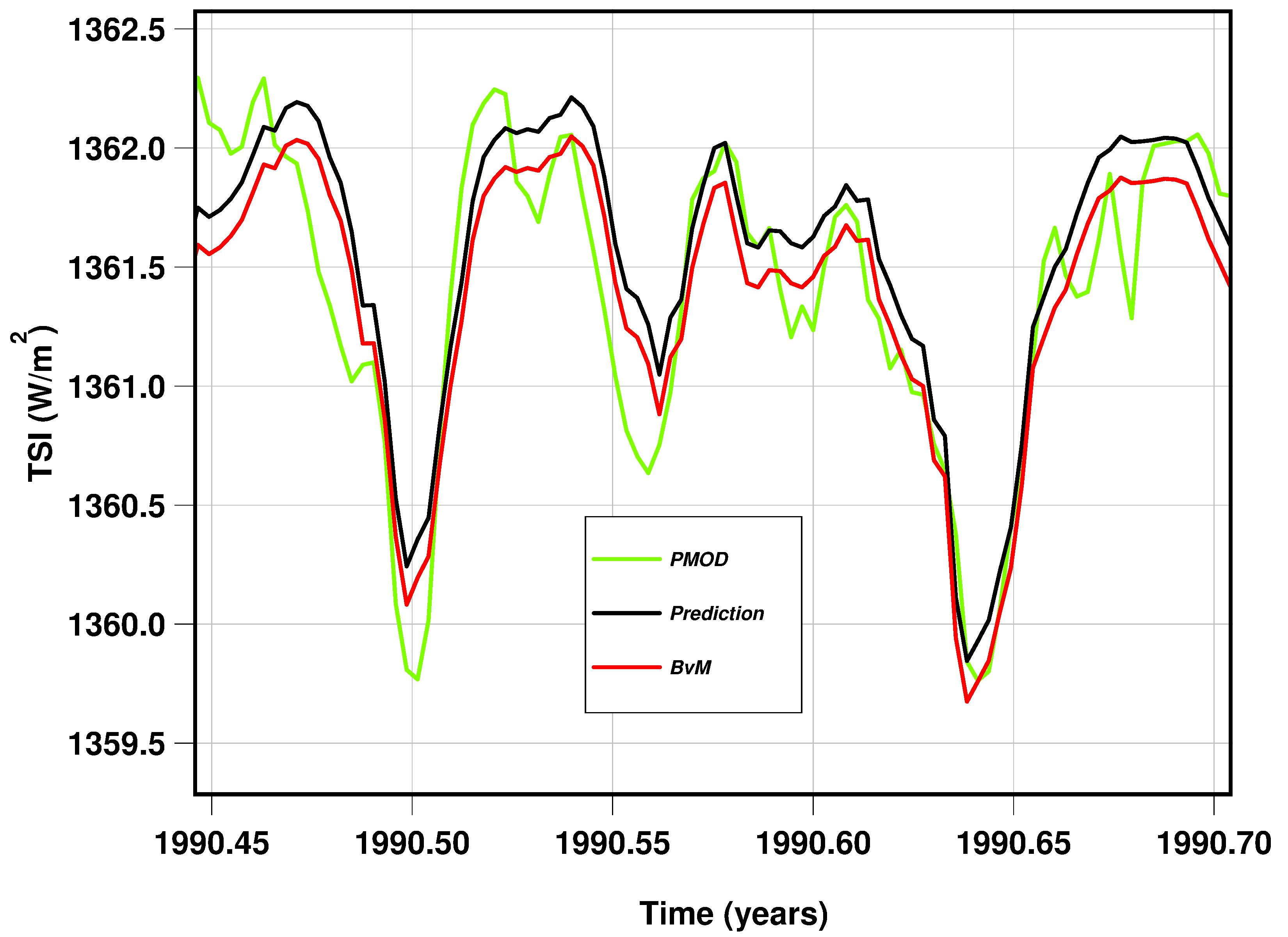}}
\caption{Fragment of the simple prediction from PSI to PMOD together with the BvM version and the target PMOD data.} 
\label{fig008}
\end{figure}

\begin{figure} 
\centerline{\includegraphics[width=0.5\textwidth,clip=]{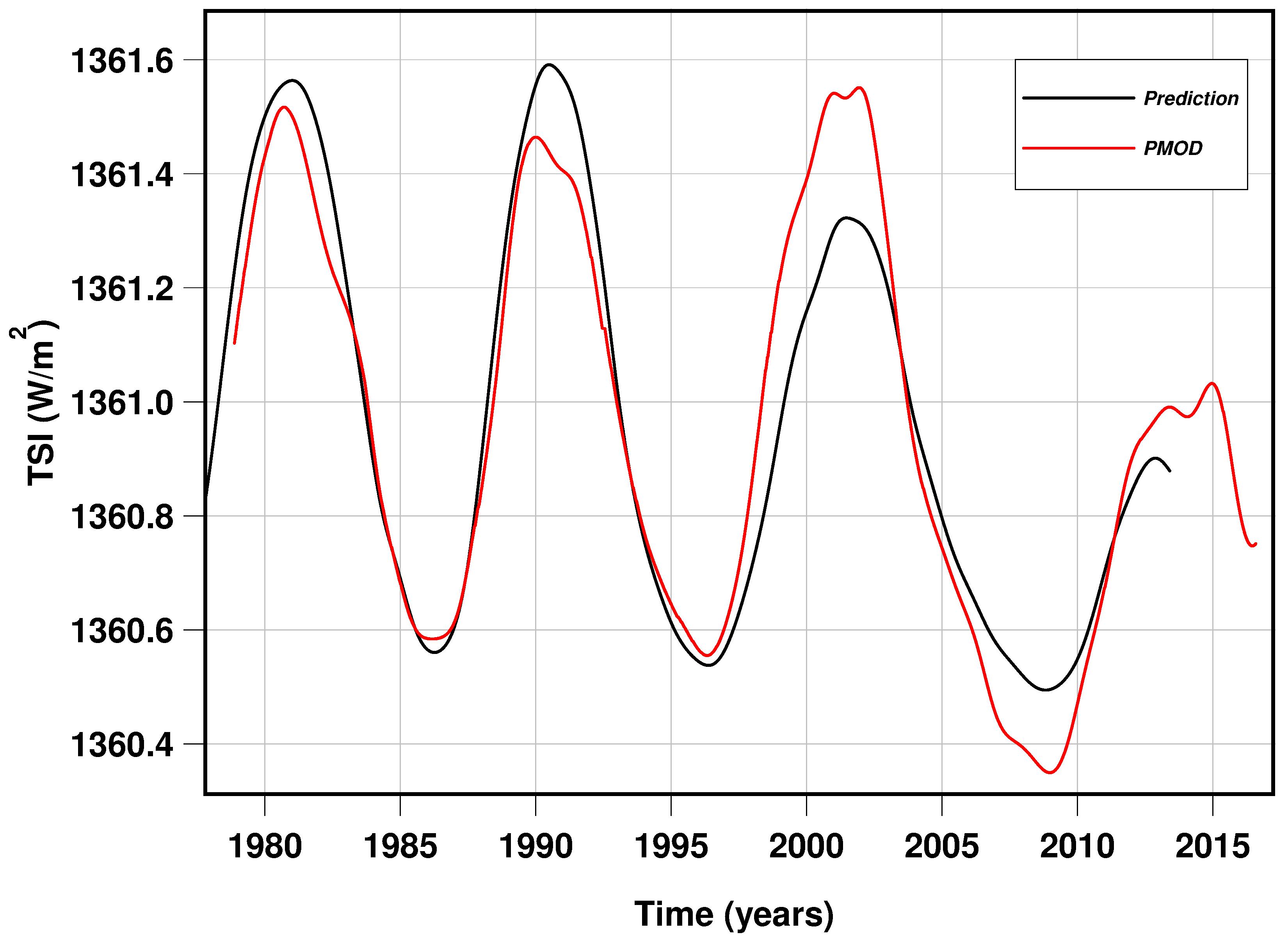}}
\caption{LF parts of the simplest prediction using PSI as the proxy and PMOD as the target. 
} 
\label{fig009}
\end{figure}

In Fig.~\ref{fig008} short fragments of 
the target data set PMOD, of the predicted curve, and of the BvM-corrected curve
are plotted to show that the PSI curve, if mapped properly, can
model PMOD as target. 
The part of the residual differences originates from the brightening events 
that are not accounted for in sunspot statistics and/or do not correlate with it.  
Another difference is due to the unavoidable modeling errors.   
In Fig.~\ref{fig009} we plot the LF parts
of our reconstruction based on the simple model (Eq.~\ref{simplemod}) and
the target PMOD.  The difference between these curves is only
the BvM correction we applied to our solution
to form an idea how precisely short timescale fluctuations in the TSI curve
can be modeled by the raw PSI data.

To recapitulate this part of the analysis: the raw PSI data used as a
straightforward blanketing model combined with correct LF part
helps us to achieve regression modeling precision at the level of $R_c=0.887$ (PSI vs. PMOD) or $R_c=0.898$ (PSI vs. ACRIM),
which is indeed quite high. For the hindcasting problem this is important. 
We do not have proper data to estimate the daily brightening component for
the historical data, but this may not be very important. For
the proper LF part, the raw PSI data as a regression component are very useful.  
In the next section we further improve on this by using
additional PSI (or other proxy) -based regression components.

\begin{table}
\caption{Optimal $W$ values (breakpoints) for the simplest models.}
\begin{center}
\begin{tabular}{lrrr}
  & PMOD & RMIB & ACRIM \\
\hline
PSI &   791.9 &   824.6 &   693.7  \\ 
SA &   807.3 &   {\bf 825.8} &   664.7  \\ 
SN &   734.9 &   628.7 &   \iti{ 472.7}  \\ 
\hline
\end{tabular}

\label{table005}
\end{center}
\end{table}

\begin{table}
  \caption{Prediction vs. target correlations for the simplest regression model and for the BvM variant of it.}
\begin{center}
\begin{tabular}{lrrr}
  & PMOD & RMIB & ACRIM  \\
  \hline
  Simplest &&&  \\
  scheme & & &  \\
\hline
PSI & {\bf     0.860 }  &     0.818 &     0.749  \\ 
SA &     0.798 &     0.744 &     0.689  \\ 
SN  &     0.711 &     0.649 & \iti{     0.644 }  \\ 
\hline
BvM & & &  \\
\hline
PSI &     0.887 &     0.885 & {\bf     0.898 }   \\ 
SA &     0.831 &     0.833 &     0.865 \\ 
SN & \iti{     0.741 }  &     0.745 &     0.813 \\ 
\end{tabular}

\label{table006}
\end{center}
\end{table}

\subsection{Application of the simple scheme}

We then applied the simple regression scheme described by Eq.~(\ref{simplemod}) to
all the different sunspot-tracing proxy-target pairs. In Table~\ref{table005} we list the
optimal breakpoint $W$ values for different pairs. These values
tend to be rather similar between all the targets. The correlation
levels achieved by using the simple prediction scheme are given in
Table~\ref{table006}.  In this table, we present two types of
results: in the first part (simplest scheme), we list the correlation values
between the simplest model prediction and real data.  In the
second part (BvM), we display correlations obtained by our BvM scheme
with the backbone substitution.

\begin{figure*} 
\includegraphics[width=0.5\textwidth,clip=]{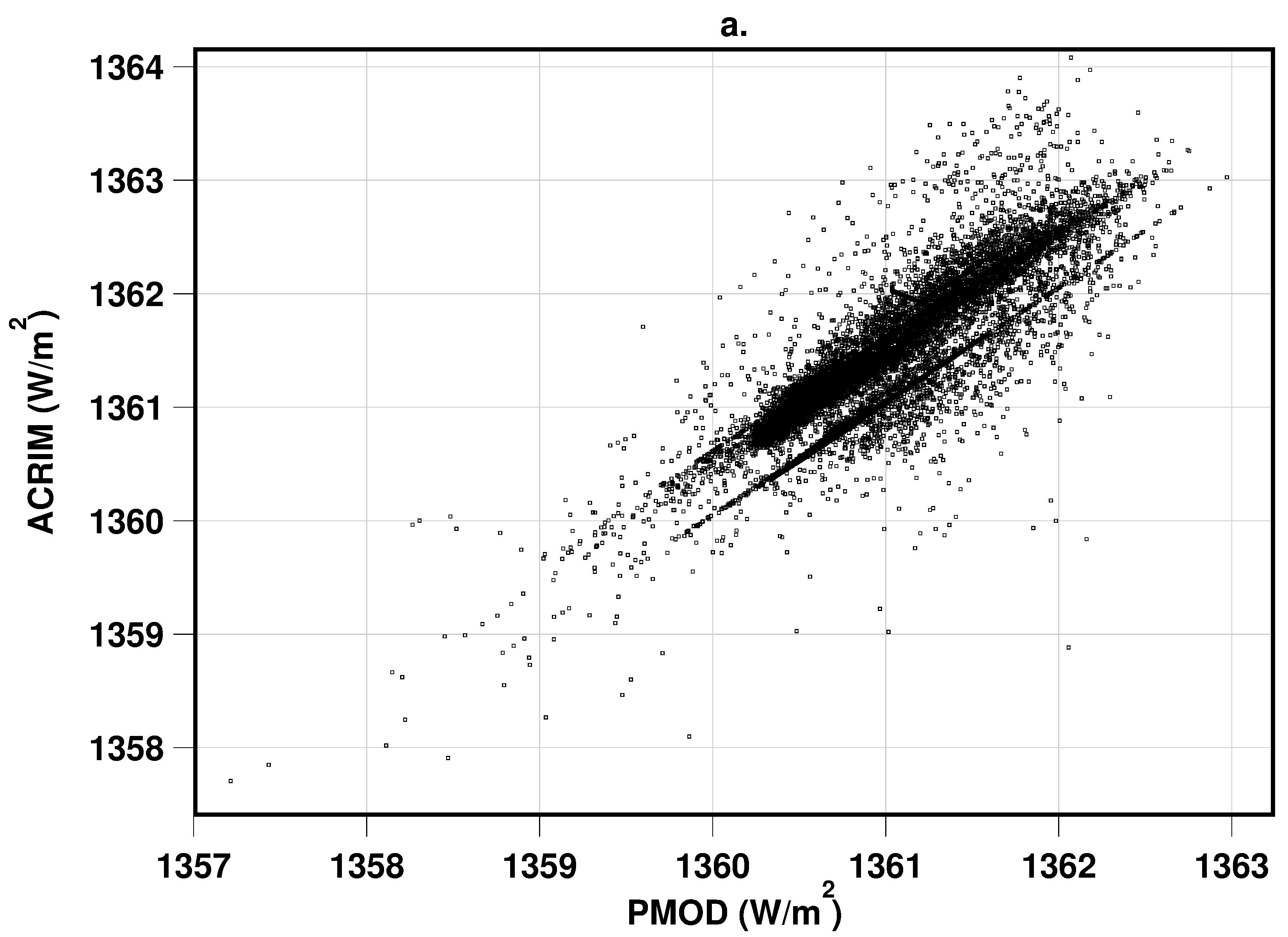}
\includegraphics[width=0.5\textwidth,clip=]{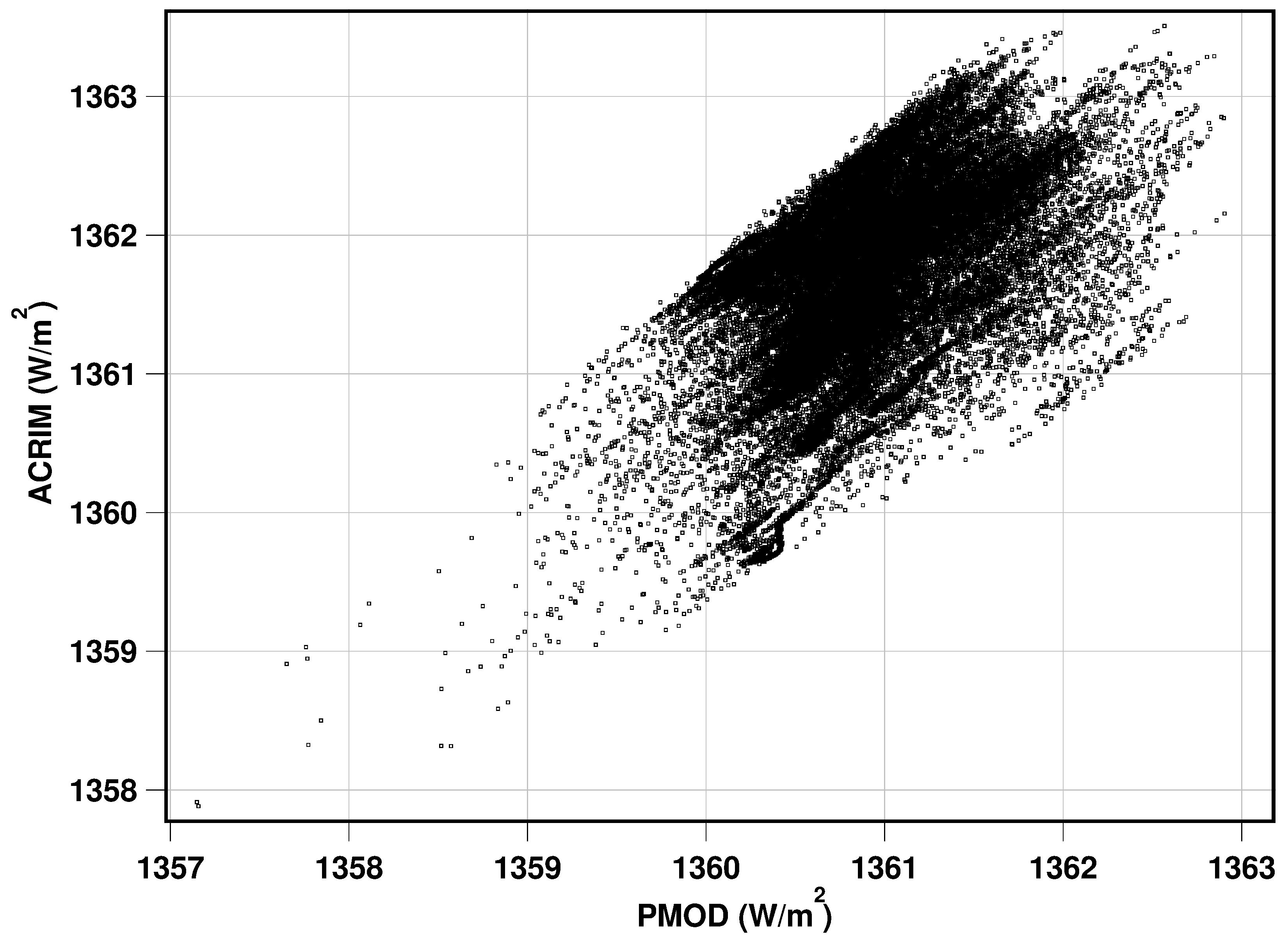}
\includegraphics[width=0.5\textwidth,clip=]{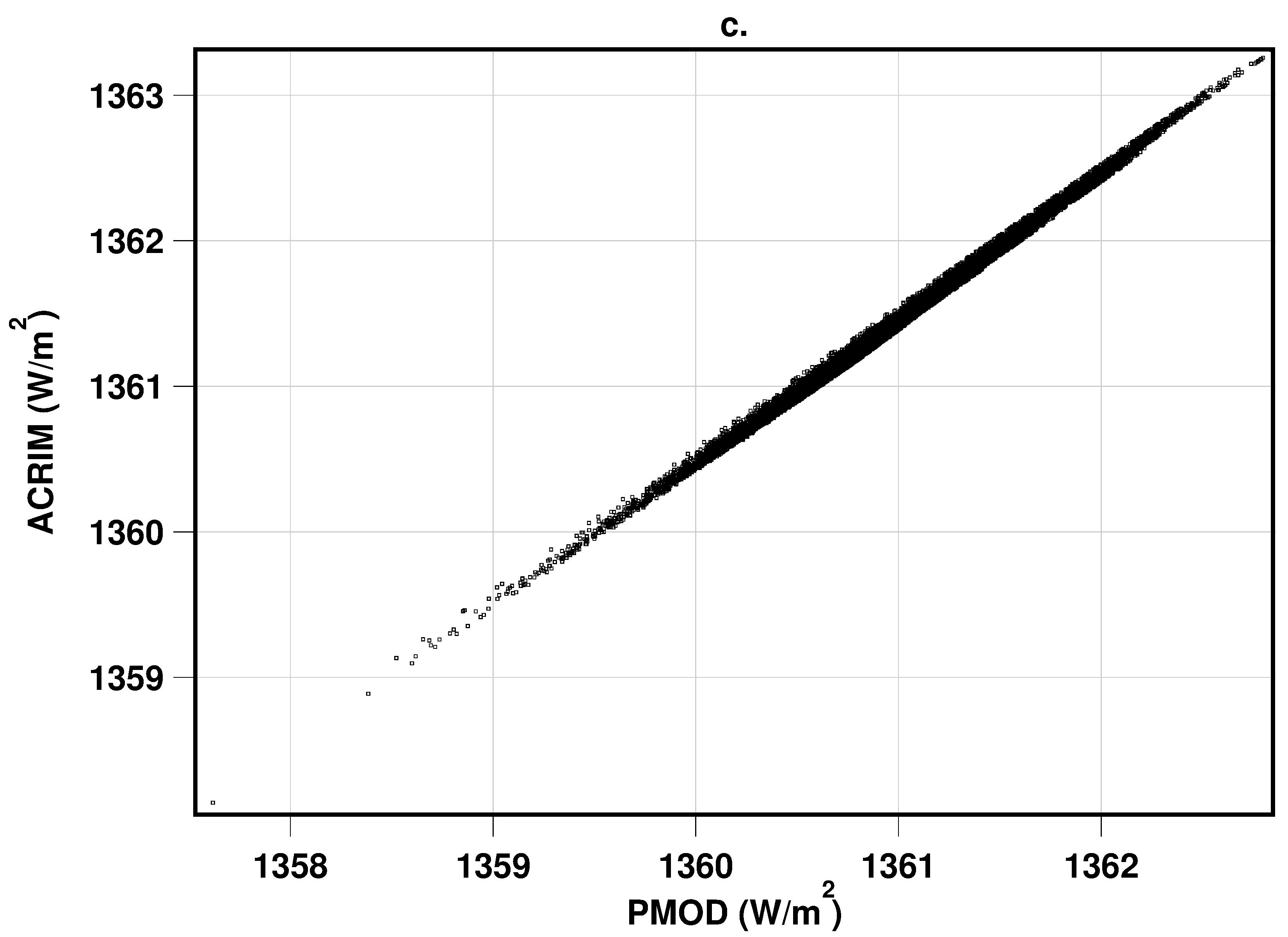}
\includegraphics[width=0.5\textwidth,clip=]{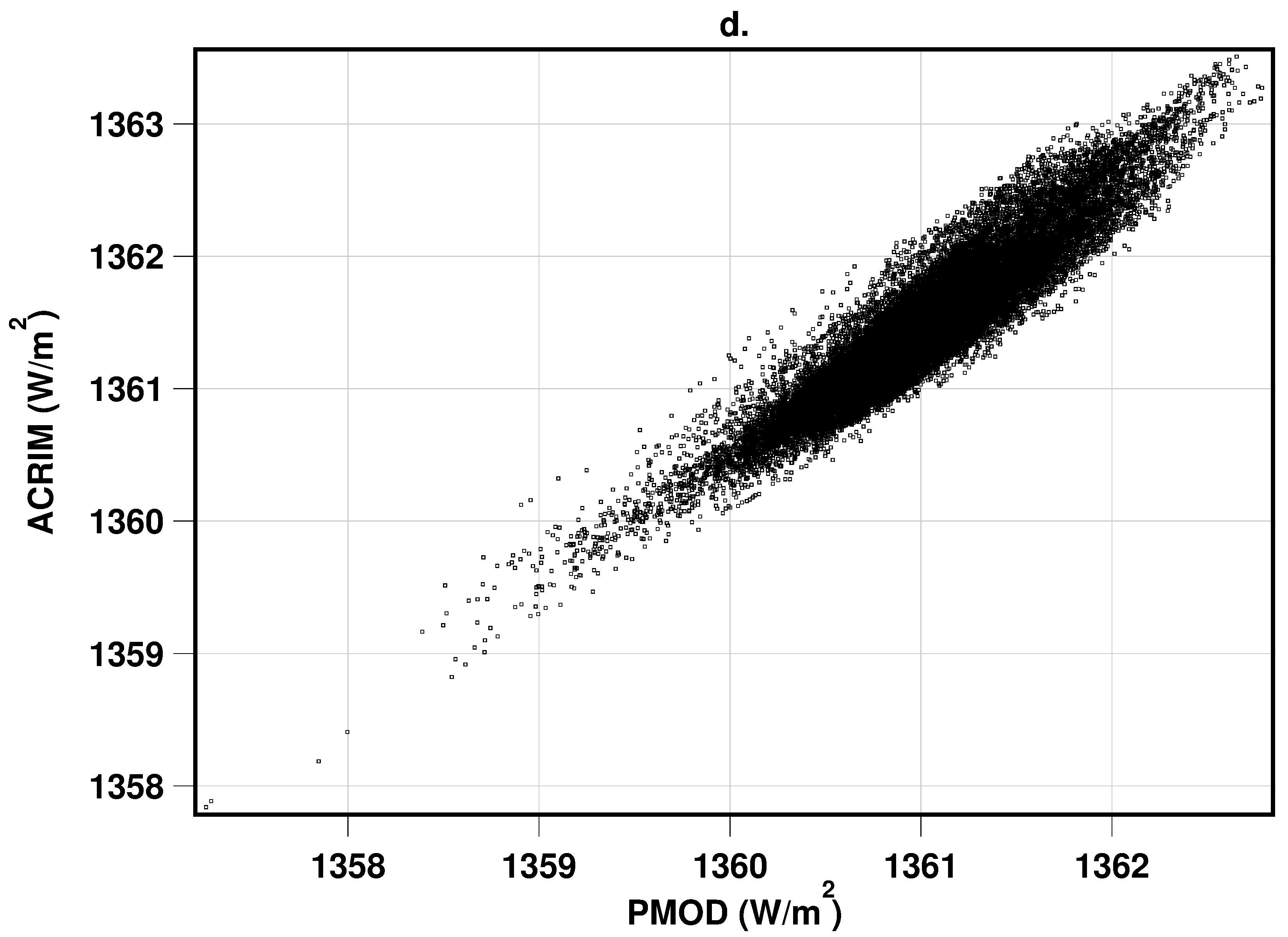}
\caption{Correlations between PMOD and ACRIM and regression models based on them (using PSI as the proxy): 
  a. cross plot of the original target data sets PMOD vs. ACRIM ($R_c = 0.861$),
  b. cross plot for the solution without restrictions ($R_c = 0.512$, Fig.~\ref{fig013}, see Appendix~\ref{appa}.);
  c. cross plot based on the simplest models ($R_c=0.998$),
  d. cross plot between restricted multicomponent models ($R_c = 0.941$).} 
\label{fig010}
\end{figure*}

We computed TSI hindcasts for every proxy-target pair
and compared them by computing correlation coefficients between
the different solutions. 
The four-dimensional correlation matrix
obtained for different proxy-target pairs is given in
Table~\ref{table17} (Appendix~\ref{appb}). 
Different solutions are inherently
quite consistent. This is also demonstrated in Fig.~\ref{fig010}
c. for the particular combination PSI-PMOD vs. PSI-ACRIM.  The
similarity of almost $100\%$  between the two hindcasts results from the
fact that all information about the targets is compressed into
only four
estimated coefficients: the linear parameters $a_0$, $a_1$, and $a_2,$ and
the nonlinear smoothing parameter $W$. If we take into account that the
correlation computation itself balances out two of them (mean level
and dispersion), then we are left with only two parameters.

The correlations between targets and the prediction fragments are
significantly more scattered for different proxy-target pairs 
(Table~\ref{table006}). Part of this scatter is a result of our
rather trivial modeling method (raw proxy as a model for
blanketing). The other part is due to the different predictive
capacities of the proxies (e.g., PSI vs. SN). Finally, the large spread of different target values also influences
the prediction and hindcasting outcomes.
 
Nevertheless, the correlation levels achieved using this simplest
scheme (e.g., $0.860$ for the PSI-PMOD pair) are rather high. 
The corresponding hindcast TSI curves can well be considered as simplest
solutions for the hindcasting problems.
The much more complex
models described next increase the level of proxy-target correlations,
but also loose some robustness and stability.

\section{Multicomponent regression models}\label{s:multimod}

The simplest regression models described above only take
two important features of FDCs into account (see Fig.~\ref{fig007}):
the central LF peak, and the wide anticorrelation depression. 
To account for narrow peaks in this spectrum, we need to add components to our modeling scheme.
This is a far from trivial task. 
Of the large set of trial components, only a small
number are useful: they help to model the target and are stationary enough to have a predictive capacity.
This forced us to include various selection and restriction steps  in our final hindcasting algorithm.

\subsection{Model building} 

The simple regression scheme can be improved by adding some new
components to the simplest model:
\begin{equation}
C(t)=a_0+a_1 E[0,0](t)+a_2 E[W,0](t)+\sum_{l=3}^L a_l E_l[\dots](t),
\label{fullmod}
\end{equation}
where $E_l[\dots]$ are different transforms of the proxy data.
Using added components, we tried to model finer details, as we
show in Fig.~\ref{fig007}.
For instance, in the region of the rather broad frequency offset band of
strong anticorrelations, some peaks with significantly weaker
anticorrelation/almost complete uncorrelation can be found  at approximately the frequency offsets corresponding to solar rotation.

To build new trial components, we filtered the proxy data set with
different bandpass filters, allowing the width, $W$, and the
frequency offset parameter, $O$, to vary. We also applied different envelope
construction modes.

The full multidimensional least-squares optimization over a large set of parameters
and envelope-building modes is very time consuming. We therefore
used a
suboptimal modeling method, the so-called greedy algorithm.
In every step of this algorithm we introduce a new component from the
full library, include it in the overall regression model, and optimize it to
obtain new values for the width and offset parameters.  The values $W$ and
$O$ and the envelope mode that result in the best final correlation are
then taken as parameters of the component to be included. In principle,
this stepwise inclusion of the new components can lead us away
from the best solution.
The probability for this to happen, however, can be regarded as low because the
features in the correlation spectrum tend to stay apart, and consequently, the
sequential components are nearly orthogonal
(they do not correlate strongly with each other).
Even if some presumably incorrect component is accidentally included, the possibility remains that its effect is
decreased by some components that are added later. The overall scheme is very
similar to the cleaning approach in a frequency analysis
\citep[see][]{Roberts1987}.

For the model building the full library contains an almost complete set of 
components (in principle, the target curve can be
matched almost perfectly), but most of the low-amplitude components do
not have any predictive value: they model either noise or other
contingent aspects of the target (see, e.g., Fig.~\ref{fig013}).  For practical as well as
conceptual reasons, we therefore need to stop somewhere. In this work we chose to
include components up to the moment where the increase in
corresponding correlation level between proxy and the target is lower
than $0.001$. This choice is reasonable because, as we show below, the set of really interesting components (from the
point of view of predictive power) is rather low.

Unfortunately, as shown in the counterexample (Appendix~\ref{appa}, Tables \ref{table008}-\ref{table009}, and {Fig.~\ref{fig013})
a straightforward application of the greedy search method significantly
improves the fit between
the model curve $C(t)$ and target $Y(t)$ (for the PSI--PMOD pair $R_c=0.906$, for the PSI--ACRIM pair $R_c=0.918$),
but it produces very unstable solutions for the hindcasting problem.
Therefore we need to carefully consider which components are useful for the actual hindcasting.  

\subsection{Selection of regression components}

To achieve more stable proxy -- target hindcasts, we need to carefully select the regression components.
Here we applied three different methods to ensure we discard unwanted variants of the different proxy transformations:
\begin{itemize}
\item For the component-seeking procedure we preprocess input data 
to remove LF backbones from the proxies and from the targets. In this way, new
components will model only HF correlations; see Sect.~\ref{sss:Preprocessing}.
\item We check the predictive power of the new components by trial prediction within the available data, see Sect.~\ref{sss:Crossprediction}.
\item We also consider the physical plausibility of the components
by constraining certain parameter values, see Sect.~\ref{sss:Restrictions}.
\end{itemize}
All three methods are fully automatic, and no manual intervention is applied.
\subsubsection{Preprocessing}\label{sss:Preprocessing}
For every proxy--target pair under study, we first built the simplest prediction model as described above.
Then we used the optimal smoothing parameter $W$ from the simple model (see Table~\ref{table005}) to divide the input data sets into LF and HF parts.
By subtracting smoothed variants from the input data,  we essentially removed the effects of the 
LF correlations (central peak in Fig.~\ref{fig007}). The remaining HF parts were then used in the component-selection process as input data.
This preprocessing procedure allowed us to avoid a leakage effect where the LF parts influence HF parts and {\it vice versa}. 
A division into smooth and fluctuating parts such as this is often used \citep[see, e.g.,][]{Rypdal2012} but the breaking point is often chosen
arbitrarily. Here we determined it with the optimization procedure. 
When components were selected from HF parts, they were included as predictors into the final regression model to evaluate their relative strengths.

\subsubsection{Cross prediction}\label{sss:Crossprediction}

To distinguish between the simple goodness-of-fit and the prediction
potential of the components, we divided all HF parts of the input data sets
into two parts $I$ and $II$ with equal lengths in time. The first part
covered approximately cycles 21-22 and the second part cycles 23-24. In
every new component-seeking step we then evaluated a fit criterion
in the following way. For every possible parameter combination, we
built two separate models: for the first, $I$, and for the second
part, $II$, of the data.  Then we used these models to predict or
hindcast the other one. In this way, we obtained two correlation values,
$R_c^{I \to II}(W,O,T)$ and $R_c^{II \to I}(W,O,T),$ where $T$ denotes
the particular type for the envelope (upper, lower, or simple smoothing,
detrended or not). It is important that the model parameters in this
procedure were computed using one part of the data but the correlation is
measured between the predicted values and the other part.  For each
parameter set, we then took the minimum value for the two correlations and
maximized it to derive proper parameter values. Formally, we chose
\begin{equation}
R_c=\max_{W,O,T} \min (R_c^{I \to II}(W,O,T),R_c^{II \to I}(W,O,T)),
\end{equation}
as our final correlation estimate for a particular component. In this
way, we selected components that may be ill-suited for
detailed modeling, but perform well in the context of predictions.

\subsubsection{Search domain restrictions}\label{sss:Restrictions}
There is an additional method to cull components that can lead
to incorrect predictions. \citet{Pelt2010} showed that the magnetic
activity on the solar surface has a certain ``memory'' that is
somewhere around 7-15 solar rotations. 
This means that modes whose
wavelengths are longer than a given value can describe accidental correlations,
not systematic ones. Therefore it is reasonable to restrict the
parameter range for bandpass-filter offsets to values $O<500$
days. We assume that dependencies with longer wavelength correlations
are previously absorbed into the LF components or do not contribute as
potential sources of predictive power.

On the other hand, it is also reasonable to restrict the $W$
parameter. Values of $W$ that are too high result in bandpass filtering with
very narrow bands, and consequently, the corresponding envelopes are prone
to fluctuations (the spectral information for these modes comes
from only small set of the Fourier-transformed data values).  There is no
good quantitative method to derive this limit from the physical
principles, but from the wide range of trial computations with
different input data sets, we found that the limit $W<2000$ can well
be used as a reliable approximation.

By combining cross prediction with the restrictions for the model
parameters, we now have a method to model HF parts of our targets and
then to hindcast past TSI values. It is quite clear that this
method will provide a lower overall fit quality than the two
examples given in Appendix~\ref{appa}, but it is expected that the hindcast quality
will increase.

\subsection{Computed models}\label{ss:compmod}

Our final regression model for the hindcasting of a target using a proxy
then consists of (the parameters that are to be estimated are
added in parentheses):
\begin{itemize}
\item constant level ($a_0$),
\item the proxy curve itself ($a_1$),
\item an optimal LF component ($a_2$), 
\item set of components from the HF analysis ($a_3,\dots$).
\end{itemize}
The typical component-seeking process is illustrated in Table~\ref{table010} for the
PSI-PMOD input data pair. First the values of the determined component parameters are given, and then
the estimated regression coefficients. In the final two columns the convergence process
of the iterations is illustrated by listing at each step the
correlation level that has been achieved and its increment.
Similar data for PSI-ACRIM pair are presented in Table~\ref{table011}.
These tables show that all different solutions
contain components with offset parameter values around $O=27$d,
which corresponds to the solar rotation, as
Fig.~\ref{fig007} indicated.
Similarly to the rotation signal, the components lie in the interval
from $O=9$ to $O=12$ in the different solutions. These components appear
because the method tries to
take the features in the transient part (from anticorrelation
to decorrelation) of the FDC spectrum into account. The remaining components describe
low-frequency features and differ more from one method to the
other. In principle, they absorb more contingent features of the
variability, and consequently, they depend more strongly on the method
that is used. Certainly the solutions based on cross prediction, even if they are
slightly less correlated with the learning sets (targets), must be
taken more seriously.

In Fig.~\ref{fig011} we plot multicomponent hindcasting solutions for the pairs PSI-PMOD and PSI-ACRIM together
with the LF components of the target curves PMOD and ACRIM. The statistical correlation between
the solutions is rather high ($R_c=0.941$), but the curves differ somewhat. Subjectively, we would prefer
the first solution, but according to the approach taken in this paper, we treat all solutions equally. This plot can be also compared with Fig.~\ref{fig013}, where unconstrained modeling
results are depicted in a similar format. The introduced selection procedures for model components
allow more stable solutions.     

The obtained correlation levels for the entire set of proxy-target
combinations is presented in Table~\ref{table012}.
In the first group of
the table (HF prediction), we list the correlations achieved by modeling the HF
parts of the corresponding proxy and target. The second
group (Prediction) consists of the final correlations of the model, where
both the component smoothed by optimal values of $W$ and
components found from HF analysis are included.  And finally, as with simplest models, the third group displays correlations
for the mixed models, where LF part is substituted by backbone
from target (BvM method).

\begin{table*}
\caption{Modeling PMOD data using PSI as a proxy. Regression components and iteration progress.}
\begin{center}
\begin{tabular}{llrrrrr}
Coef. & Type & $W$ & $1/O$ & Value & $R_c$ & $\Delta R_c$ \\
\hline

$a_{0}$ & 1.0 & - & - &    1361 & & \\
$a_{1}$ & $E$ & 0 & 0 &      -0.8644 & & \\
$a_{2}$ & $E^{ }$ &     791.9 &  -  &      11.81  & 0.8604 & \\
$a_{3}$ & $E^{ }$ &     374.4 &      27.16 &      14.98 &       0.8732 &       0.0128 \\
$a_{4}$ & $E^{ }$ &      11.10 &      10.60 &     -14.91 &       0.8794 &       0.0062 \\
$a_{5}$ & $E_{d}^{-}$ &       7.271 &      11.22 &       3.600 &       0.8822 &       0.0029 \\
$a_{6}$ & $E_{d}^{-}$ &      78.30 &     186.5 &       6.007 &       0.8846 &       0.0023\\
$a_{7}$ & $E_{d}^{-}$ &     877.8 &     150.5 &      -8.633 &       0.8863 &       0.0017\\
$a_{8}$ & $E^{-}$ &     236.7 &     606.5 &      -7.428 &       0.8880 &       0.0017 \\
$a_{9}$ & $E^{+}$ &      96.97 &     338.1&     698.8 &       0.8893 &       0.0014 \\
$a_{10}$ & $E^{+}$ &      96.83 &     338.1 &    -695.7 &       0.8913 &       0.0019\\
$a_{11}$ & $E^{-}$ &     158.6 &       9.725 &     -22.18 &       0.8925 &       0.0012\\
$a_{12}$ & $E^{ }$ &     235.9 &       8.952 &     -14.38 &       0.8933 &       0.0009\\

\hline
\end{tabular}
\label{table010}
\end{center}
\end{table*}

\begin{table*}
\caption{Modeling ACRIM data using PSI as a proxy. Regression components and iteration progress.}
\begin{center}
\begin{tabular}{llcccrr}
Coef. & Type & $W$ & $1/O$ & Value & $R_c$ & $\Delta R_c$ \\
\hline
$a_{0}$ & 1.0 & - & - &    1361                                       & & \\
$a_{1}$ & $E$ & 0 & 0 &       1.016                                       & & \\
$a_{2}$ & $E^{ }$ &     693.7 &  -  &      10.06                      & 0.7452 & \\
$a_{3}$ & $E^{ }$ &     368.4 &      27.34 &      15.10             & 0.7559 & 0.0107 \\
$a_{4}$ & $E^{+}$ &     224.0 &     100.5 &     -25.71             & 0.7649 & 0.0089 \\
$a_{5}$ & $E^{+}$ &     794.3 &     138.1 &      52.45             & 0.7712 & 0.0064 \\
$a_{6}$ & $E^{ }$ &      12.33 &      10.90 &     -21.76             & 0.7766 & 0.0054 \\
$a_{7}$ & $E^{ }$ &    1034 &  -  &      87.41                      & 0.7808 & 0.0042 \\
$a_{8}$ & $E^{+}$ &     716.7 &       9.057 &    -139.9             & 0.7871 & 0.0063 \\
$a_{9}$ & $E_{d}^{+}$ &      76.92 &     118.0 &       6.385         & 0.7899 & 0.0028 \\
$a_{10}$ & $E_{d}^{-}$ &      25.51 &      39.20 &       3.278        & 0.7915 & 0.0016 \\
$a_{11}$ & $E^{-}$ &     107.3 &       9.643 &     -24.05            & 0.7934 & 0.0019 \\
$a_{12}$ & $E_{d}^{-}$ &     100.3 &      98.60 &       7.281        & 0.7946 & 0.0012 \\
$a_{13}$ & $E^{ }$ &     204.0 &     101.1 &       8.631            & 0.7956 & 0.0009 \\
\hline
\end{tabular}
\label{table011}
\end{center}
\end{table*}

\begin{table}
\caption{Achieved correlation levels for the multicomponent regression model.}
\begin{center}
\begin{tabular}{lrrr}
  & PMOD & RMIB & ACRIM  \\
\hline
HF prediction & & &  \\  
\hline
PSI &     {\bf 0.831} &  0.816   &     0.769 \\ 
SA &     0.690 &     0.707 &     0.657 \\ 
SN &     0.488 &     0.469 & \iti{     0.400 }  \\ 
\hline
Prediction & & &  \\
\hline
PSI & {\bf     0.893 }  &     0.848 &     0.796 \\ 
SA &     0.842 &     0.811 &     0.756 \\ 
SN &     0.757 &     0.686 & \iti{     0.674 }  \\ 
\hline
BvM method & & &  \\
\hline
PSI & {\bf     0.914 }  &     0.909 &     0.908 \\ 
SA &     0.869 &     0.862 &     0.874 \\ 
SN &     0.783 & \iti{     0.769 }  &     0.820 \\ 
\hline
\end{tabular}
\label{table012}
\end{center}
\end{table}

Table~\ref{table012} and Fig.~\ref{fig006} c show that the correlation levels achieved using
the multicomponent regression models lie between the levels of the simple
prediction schemes and models to which components were added without any
restrictions (see Appendix~\ref{appa}).  For PSI-PMOD pair the corresponding levels are $0.860$
(the simplest method), $0.893$ (multicomponent method), and $0.906$ (unrestricted method). For
the PSI-ACRIM pair the respective levels are $0.749$, $0.796$ and $0.918$.

\begin{figure} 
\centerline{\includegraphics[width=0.5\textwidth,clip=]{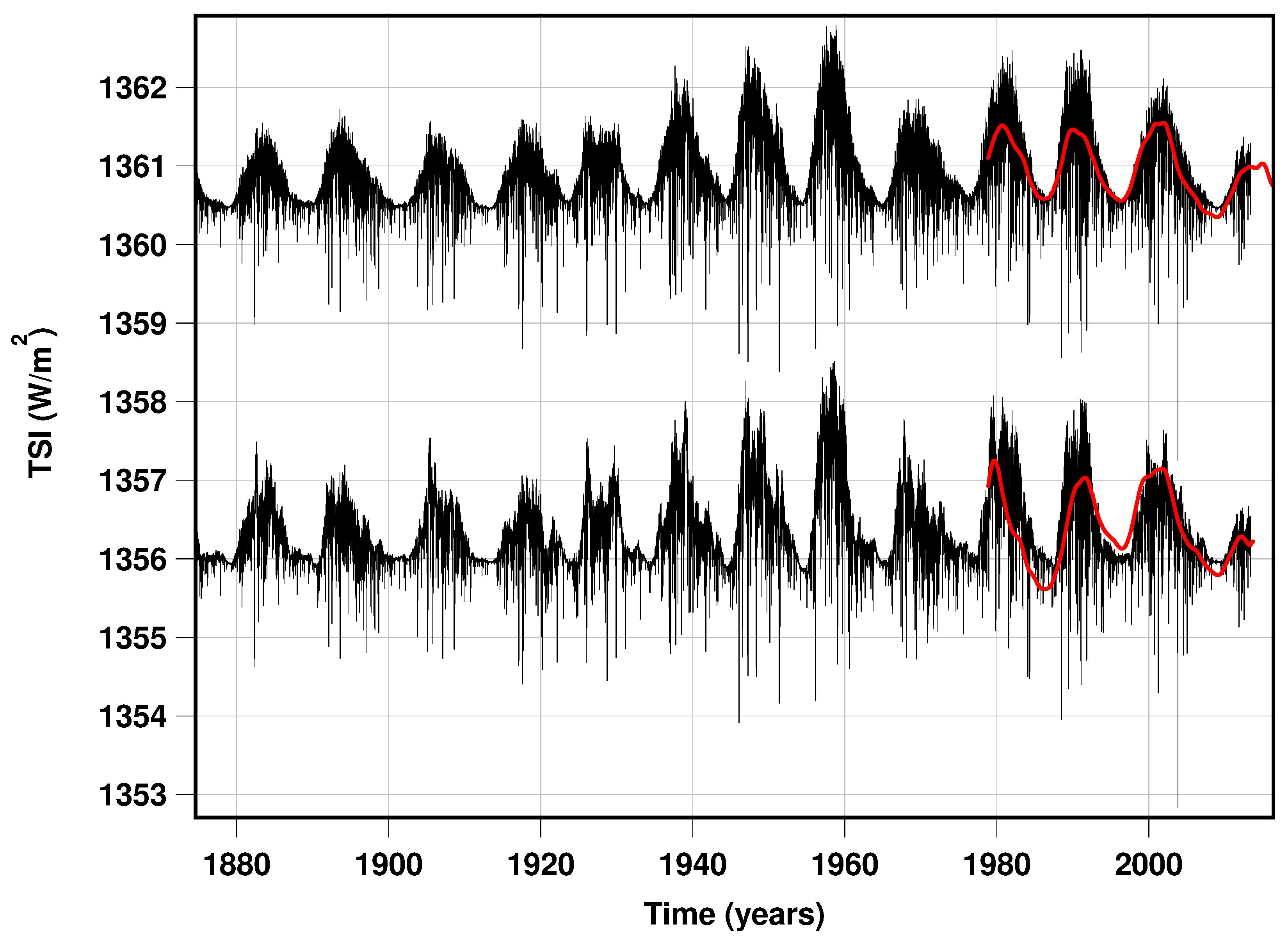}}
\caption{Hindcasting based on multicomponent models. 
PSI to PMOD mapping (upper curve) and PSI to ACRIM (shifted down by five units; lower curve). In red we plot the low-frequency
backbones of the target curves.
The cross correlation between the two hindcast curves is $R_c = 0.941$ (see Fig.~\ref{fig010}d). 
} 
\label{fig011}
\end{figure} 
Probably the most striking result is the very similar correlation level
that is achieved with the BvM method (see
Table~\ref{table012}). From this it follows that all the
three targets are practically equivalent when we consider their HF
behavior (blanketing by sunspots and short-term enhancements). All
the problems and differences between the targets originate from their
LF behavior.

The actual predictions (Table~\ref{table012}, the Prediction part) differ more significantly. This is also illustrated in
Fig.~\ref{fig010} d, where the cross plot between the PSI-PMOD
prediction and the PSI-ACRIM prediction is displayed. The correlation
level is certainly better than with the unrestricted model
solution (b), but this is far from being the case with the simple model (c).

In Fig.~\ref{fig012} we compare the prediction errors for the final
multicomponent solution (upper panel) and the BvM variant of it (lower panel)
that were computed for the PSI-PMOD proxy-target pair. The day-to-day
error values in the plot are somewhat misleading because of the low
resolution of the plot. Differences (errors) between monthly and yearly
averages show the expected level of precision for the applications
where such averages are used as an input. Unfortunately, the plot
also demonstrates the level of ambivalence that is due to the main
controversy, that is, to the varying LF behavior of the targets (the BvM method
shows how the solution behaves when the LF part is
adopted from the real data).

\begin{figure*} 
\includegraphics[width=\textwidth,clip=]{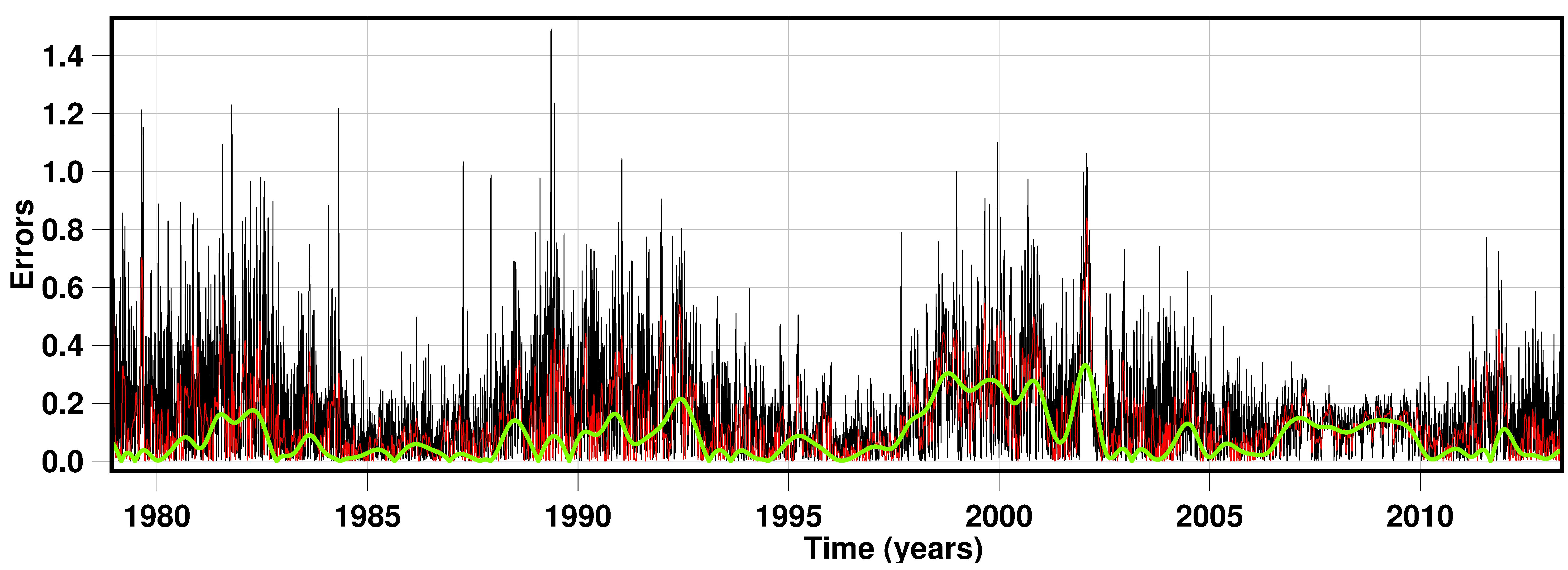}
\includegraphics[width=\textwidth,clip=]{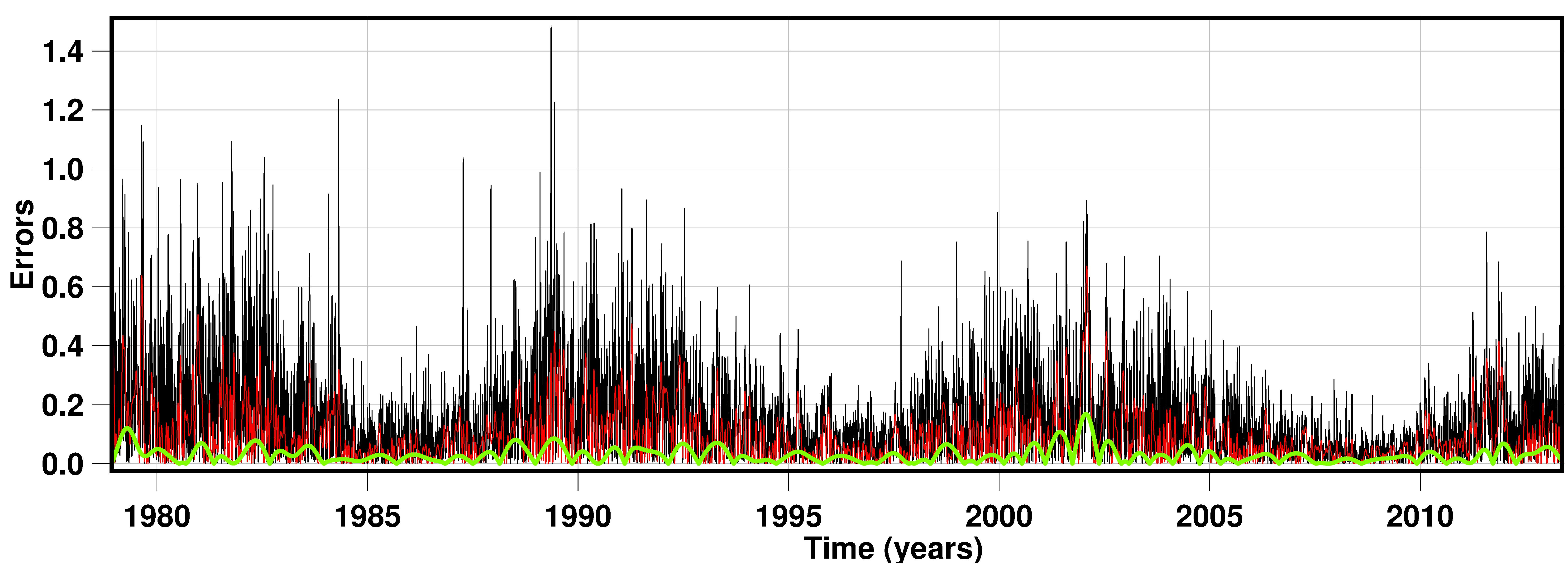}
\caption{Prediction errors (absolute differences) between predicted curves and targets for the PSI to PMOD pair (upper panel) and
corresponding BvM solution (lower panel). Black shows the daily errors, red the errors for filter curves smoothed by $W=30$, and green the errors for
filter curves smoothed  by $W=365$ . The data sets are nearly 14000 points long, and consequently, the plot points essentially mark the local 
maxima of the curves (for each pixel). The scatter of the errors can also be characterized by
their standard deviations: 0.169, 0.113, and 0.076 for the upper panel and 0.156, 0.090, and 0.025 for the lower panel.}  
\label{fig012}
\end{figure*}

\subsection{Postprocessing}
After the hindcasts are built with different simple and multicomponent regression models,
some refinements can be applied to the final data products.    

The scatter of the correlation levels between PMOD and PSI
(Fig.~\ref{fig007}) strongly oscillates at higher
frequencies. When we compute FDCs using a wider passband, for instance,
setting $W=30$ instead of $W=2000$, we obtain a much smoother curve.
From Fig.~\ref{fig007} we can see that the average level of
correlations reaches the zero level only asymptotically. This means that we
have certain minor correlations even for the very high
frequencies. These correlations are due to the very sharp peaks in the
time domain that have wide images in Fourier domain  (this is true
for targets as well as for proxies). However, because the statistical
fluctuations are of general origin, they are practically useless. Even if
we try to add many high-frequency modeling
components, we end at certain overfit situation. Consequently, it is
reasonable to cut out the high-frequency region from our prediction
scheme. Figure~\ref{fig007} shows that a frequency of
$0.2$ cycles per day is a reasonable limit. The effect of this minor
smoothing with a $W=5$ Gaussian filter is not very dramatic. For instance,
in the case of the correlation value $R_c=0.893$ between the
PSI vs. PMOD
solution and the original PMOD (see Table~\ref{table012}, column
Prediction), the correlation after smoothing both of the curves attains
the level $0.9022$. Therefore it
is reasonable from a practical viewpoint  to
mildly smooth the final data products. Even if we claim that the obtained
results are given as day-to-day data, the other users must be aware
that some minor details are not presented in hindcasts, especially
that the
sharpest peaks are somewhat rounded at their extrema. Fortunately, most of the applications need only data on much coarser grids, and then these
minor details are certainly averaged out. 

Obviously, our methods do not help at all in the context of
estimating the absolute level of the TSI even if this is an important aspect of the Sun-Earth relations 
\citep[see][]{Kopp2011}. However, for presentation purposes, we need a
certain fixed point to tie  the different time series together. There
is one obvious candidate to clearly show the overall level of
the TSI: the cycle 23-24 minimum ($1360.85 Wm^{-2}$ at 15 October 2008
\citep[see][]{White2011}).  However, this leads to a wide
spread of targets for exactly the time when the Sun worked as usual. We therefore chose as a compromise another minimum between cycles 22-23 (1361.11 at
1996.465, see Fig.~\ref{fig005}) using a similar smoothing method
as in the original paper \citep{White2011}.

All the results are available in the form of simple text files that can be downloaded\footnote{\url{http://www.aai.ee/~pelt/TSI}}.
The predicted time series are all shifted to a common mean level and are mildly smoothed, as described in the last two
subsections. The input data sets and modeling methods for every prediction can be read off from their file names. 

\section{Discussion}\label{s:disc}
From the statistical point of view, the TSI consists
of three components: variability that originates from a straightforward
blanketing effect
due to the sunspots, variability that statistically correlates with
sunspot occurrences (or their areas), and residual variability.  The
last component contains long time trends of overall variability and
short time features that occur randomly and are not statistically
correlated with sunspot occurrences. Part of the residuals is also due
to the nonlinear correlations, which are more complex than the correlations that are
accounted for by modeling using smoothing and envelopes.

The method of FDCs allowed us to quantify these observations and to
localize different aspects of variability in the frequency spectrum
\citep[for a more traditional description of the time variability, see
the review by][]{Solanki2013}. From the physical point of view, we
can classify the solar variability in even more detail. Every spatial
disturbance in luminosity (regardless of spectral region) also
translates into the time domain due to rotation.  Evidently, this
translation is far from simple: The different activity tracers
are likely to have phase shifts and undergo phase mixing. Nor can the
translation be expected to occur over one discrete frequency, as
the rotation velocity depends on both latitude and radius of the Sun.
In addition, we can have certain secular changes that result from the
particular way the solar dynamo operates.  The recent DNS model of
\citet{Millennium}, which covers roughly 80 simulated solar cycles, suggest
that prominent secular changes can indeed occur as a result of the
existence of multiple dynamo modes with cycles of varying
frequency. Even though the cycles seem to be rather regular in
frequency, their interference can cause abrupt phenomena that
are reminiscent
of the Maunder minimum in addition to a smooth secular component.

As a result of our smoothing experiments with varying window width, we
found that the strategically most important breaking point is
localized at the frequency that corresponds to a period of $\approx 750$
days. We used this dividing line to build LF and HF components of the
proxies and targets. When comparing the LF parts of the proxies (PSI, SA,
and SN), we can conclude that they are rather similar and
essentially contain
the same information. These curves characterize the slowly
changing mean level of solar activity, which shows itself through the
statistics of sunspot occurrences. The similarity of the LF component of the
proxies is a hindrance for the use of multiple regression TSI models
\citep[see, e.g.,][]{Zhao2012}, where different proxies are
combined to obtain TSI estimates.
The models can still be improved, however, by increasing the accuracy
of the HF part description.

The LF parts of the target data sets (PMOD, RMIB, and ACRIM)
differ more from each other. The differences between them originate
from certain instrumental, data processing, and other method-related
problems \citep{Kopp2014}.
For TSI hindcasting, the LF variability is the strongest hindrance
because to recover TSI values in the past, we need a good understanding
of the present-day values, which is currently unavailable.

The HF parts of the proxies differ because they are
constructed from the observed data in different ways.
The HF part is relevant for high-resolution prediction (up
to daily accuracy level). Here the proxies can be ranked by their
information content.
The best of them is PSI, which, if used in a proper regression scheme,
was shown here (and in earlier work) to describe a very large amount
of target variability. However, if we need hindcasts
extending farther back in time,
then the SN becomes useful.
How much information we loose by using these rough data can well be
deduced from the above computations.

Using the Baron von Munchausen scheme, we demonstrated that the
HF behavior
of all three targets is very similar. For instance, after backbone
substitution, the achievable correlation levels for the PSI proxy
were in the narrow interval $0.908$(ACRIM)- $0.914$(PMOD) (see
Table~\ref{table012}). This similarity is of course the result of the
essentially common origin of all the three target data sets. The main
differences among them come from different methods of fragment
stitching, not so much from rescaling or interpolations.

Probably the most unexpected result of the current analysis is the
relatively high prediction capacity of the simplest models. For instance, the simple model prediction for the PSI-PMOD pair
correlates with the target at the level $R_c=0.860$ (or, in other words,
describes a $74\%$ of variability). For 30-day running means of the
target and predicted curve, the correlation level is already
$R_c=0.924$ ($85\%$).  Consequently, if we trust any of the TSI
compilations (e.g., PMOD), then we can extend it into the past with reasonable confidence. In essence, the simplest method consists of estimating one nonlinear parameter $W$ and one ratio between
amplitudes (the LF part and the original proxy).  The other nice property of
the simplest scheme is the inherent consistency of the different solutions
(see Table~\ref{table17}).

In the simplest models we involved only two aspects of the proxy
variability: the LF part as a model for the smooth overall activity
variation, and the original proxy as an approximation of the blanketing
effect by sunspots.  There is certainly an amount of
variability that is not yet accounted for (brightness enhancements,
etc.), and therefore we can extend the simplest models by searching for more regression
components. From the wide library of the possible modes, we selected
those whose parameters lie in the physically plausible ranges,
which have the capacity to predict from one part of the series to the
another and which bring significant improvement to the final
correlation between the predicted curve and target. This procedure of
building these multicomponent models is rather time consuming and
unfortunately not very productive 
(for the particular set of proxy-target pairs). 
The final increase in correlation
levels between simple and multicomponent models is only around $3-4\%$, and
we pay for this increase with a widening spread between different
solutions. For instance, the maximal spread between different
proxy-target solutions 
(see Table~\ref{table18}) is
$0.8879-0.9836$. We can compare this with the range for simple models
$0.9957-0.9997$ (Table~\ref{table17}). 
Of course, this is expected. To calibrate the simplest model, we used only two
adjustable parameters that are computed from the targets. For multicomponent
models, the set of adjustable parameters (regression coefficients) was
significantly larger. The discrepancies between different target
curves (our main concern) are transported into the solutions for predictions
and cause them to spread out.  As we showed in the example with
unrestricted models (see Fig.~\ref{fig013}), the spread can be even
wider when we ignore physical constraints and the prediction capabilities of
the different modes.

The low degree of improvement achieved by including additional components
can also be explained by the low level of actual correlation between day-to-day values of the two components of the TSI variability: blanketing due to
the sunspots and brightening due to the faculae.  The LF behavior of
these two components can be rather similar as they result from the
occurrence and disappearance of the active regions. Nevertheless,
their HF behavior is different.

\subsection{Particular proxy-target pairs and expected precision of hindcasts}
The backprediction precision of model-based hindcasts can be estimated only very approximately. 
The computational and statistical errors
(e.g., due to the estimation of regression coefficients) are
significantly lower than the overall spread of different solutions, so
that it is not even reasonable to tabulate them.  The entire
spread of
the solutions originates from different input data sets and from
differences in regression model compositions.

There are some components that are quite similar for all the $3\times
3= 9$ models. The breakpoint parameters $W$ (Table~\ref{table005})
for different proxy-target pairs lies in the interval $472.7-825.8$
days. The smoothness of the corresponding backbones is visually
rather similar, and their differences (if relevant) must reveal
themselves in slightly different additional modes.

Another similarity between the different models occurs near the solar
rotation period. All component sets contain a mode with a period
of approximately $27$ days and with a width parameter $W\approx 400$ days. This
component suppresses the effect of the proxy in the narrow
uncorrelating band near the solar rotation (see Fig.~\ref{fig007}).

The third set of common components is around parameter values $W=10$
and $O=10$ days. These components try to filter out the high-frequency regions that are decorrelated.

The remaining components can be common for some proxy-target pairs, but can also be
lacking in other sets. They model the more contingent properties of
certain combinations. A large part of the differences between the
obtained solutions stems from these components.

We can also compute correlations between different predictions. From
the $9$ proxy-target pairs we can obtain $36$ pair-pair comparisons (Table~\ref{table18}).
The most similar pairs are PSI-PMOD and PSI-RMIB ($R_c=0.984$), and the
most different are SN-PMOD and SA-ACRIM ($R_c=0.804$). These two numbers can probably also be used as limits for very conservative (and naive)
estimates for retrodiction precision.

Of course, if there are reasons to prefer some particular proxy-target pair, the outlook is
not so bleak (see Table~\ref{table012}).
The current state of affairs is
very strongly affected by the main controversy, that is, by the discrepancy between
various targets.

However, it may be possible to obtain somewhat more precise
predictions. This comes from the other approach of TSI
approximation.
Namely, we can use SATIRE-S-type physics-based models
\citep[see][]{Ball2012} as targets in our prediction scheme. In this case, we know
the actual building blocks of the TSI approximation and can use this
information for proper prediction.  We leave this approach for the
next iteration of our research program.
   
\subsection{Effect of solar rotation}
Sun-like stars show an
unexpected damping of correlations around the solar rotation
frequencies, which may confound using our method in this field
of research.  Even if for some simple models \citep[see, e.g.,][]{Lanza2007}, we can see some
localized correlations over a certain frequency,
then for the full solar data (e.g., PSI against PMOD), there is a
quite wide peak of very weakly correlating frequencies. Put simply,
the straightforward blanketing (PSI) does not reveal itself strongly
in TSI curves (PMOD) when observed through a narrow frequency band
filter. Even if small day-to-day details coincide rather well
(Fig.~\ref{fig008}), the reshuffling of events along rotation phases
and their mixing with more random facular changes results in a strong
decorrelation. We consider this aspect of our investigations very
important and will work out more details in the next paper of this
series.

\subsection{Restrictions}
Our constructions are all based on the assumption of statistical
stationarity. However, it is well known that
the solar magnetic activity has undergone some rather abrupt states of
lower activity (grand-minima-type events). Some of these events may
be the result of the chaotic nature of the highly nonlinear physical
system \citep[see][and references therein]{Zachilas2015}
or might be caused by the interference of various dynamo modes with
different spatial distributions and symmetry properties with respect
to the equator \citep{Millennium}.

It is not ruled out that the currently available set of TSI estimates
(targets) contains a certain breakpoint when the Sun changed its normal mode and switched to a new regime. In the cross prediction scheme
above, we combined backward and forward predictions to distill variability modes that change consistently over the full time.  If the two parts differ strongly in their statistical behavior,
however, then
only a small number of variability components can be persistent enough
to allow their use away from the calibration zone. In this case, our
method is not as effective as it could be in the stationary
context. It is also well known that other Sun-like stars tend to
show rather complex long-term patterns \citep[see, e.g.,][]{Olah2012},
in which case it is not ruled out that the interval of the solar
cycles 12-23 was a calm intermezzo of regularity.

Another restriction is the time symmetry of the assumed correlations,
that is,  the variability modes are selected only using frequency slots
and no time delays are involved.  However, some processes on the Sun
can have an asymmetric statistical nature in time. For instance, the
disintegration of activity complexes is one such process. \citet{Li2010} conjectured that in the relation of sunspot activity to the TSI, a 29-day time lag is involved. In principle, it is possible
to extend our component library with modes shifted in time, but this
greatly increases the model search space.

Similarly, the proxies enter the models linearly, but it is well
known that in some cases a nonlinearity along proxy values (not in
time) can be instrumental \citep[see, e.g.,][]{Hempelmann2012}. The extension of the component library
in this direction would be computationally costly, and therefore it is beyond the scope of this study. 
Different types of (implicit) nonlinearities can be accounted for by
using neural networks approach (see, e.g., \citet{Tebabal2015}).

As we showed above, the LF components of different proxies (long and shorter) correlate so strongly that it is not reasonable to
combine many of them into a regression scheme. However, the situation
is different with the
HF parts. Here some of the proxies depend
more on blanketing and some more on brightening events. By combining
some of them into a general regression model, we can achieve better
modeling precision \citep[see for a similar
approach][]{Woods2015}. Unfortunately, all the long proxies
that are suitable for hindcasting belong to the group that only depends on
sunspot statistics, no flares or similar phenomena are accounted for. Consequently, the
FDC method with multiple proxies can only be used for more precise
interpolation and stitching of the recent data.

\subsection{Applications}
The proposed scheme of building prediction models from proxy data sets
has many additional applications.  The main application naturally is
the hindcasting or true prediction. However, we foresee some other
important applications as well.  First, the scheme can be used to
fill in gaps in some observed target data by using those parts of the
data where the proxy and target are both available to build a
prediction equation according to which the gaps are then filled based
on the proxy. 
The improved data set obtained through this procedure can iteratively be
used for further refinement of the prediction scheme until
convergence is achieved.

The second important application of the new scheme is stitching
together target data from many sources, for example, from different
satellites.  In this case, we developed a bridging scheme for fast-changing
parts of the input data using all the observation sets. Then the low-frequency smooth components can be levelled off by bending them along
the low-frequency part of the proxy curve.

In some other situations the prediction scheme can be used to reveal
secular trends or outliers in the data. We built a proxy to data
bridge using all or the most typical fragments of the
observations. Furthermore, the differences between predicted and
observed values can reveal trends or outliers.
  
\section{Conclusions}\label{s:concl}
We have introduced a rather simple computational scheme
that uses bandpass-filtered signal components as building
blocks. The practical use of these components allowed us to reveal a
set of rather interesting aspects of the proxy-target relations in the
traditional context of TSI reconstruction and hindcasting. First we
confirmed (in quantitative form) with an optimized smoothing
procedure that the most serious problem of TSI treatment is the very
large variance between the low-frequency parts of the existing TSI
composites.  Then we demonstrated a rather paradoxical feature in
proxy-target pairs, namely that the solar rotational signal is not
strongly manifested in the targets and proxies, but it is clearly and
strongly visible in the frequency-dependent correlation
spectrum. Finally, we proposed a computational scheme for
building TSI models that use the well-known proxies as input
data and are calibrated against currently available estimates of
the irradiation levels.  The introduced modeling method can be used
in different contexts: gap filling and interpolation, stitching of
observed data fragments, and in retrodiction and prediction
schemes. Hindcasts computed with the new method can be useful in
climatological models for about the past 200 years.

\begin{acknowledgements}
  We are thankful to late Olavi K\"arner, who suggested the problem to us and referee for her/his useful comments. 
  We also thank Frederick Gent, Natasha Krivova, Jyri Lehtinen, and Sami Solanki for helpful discussions. 
  This work has been supported by the Academy of Finland Centre of
  Excellence ReSoLVE (Grant No. 272157; JP, MJK, NO) and by the Estonian Research Council (Grant IUT40-1; JP).
\end{acknowledgements}

\bibliographystyle{aa}
\bibliography{paper}  

\appendix

\section{Counterexample. Unrestricted multicomponent models}\label{appa}

The multicomponent model for a proxy to TSI regression can be built using the full library of prospective components. At every step we can
include the new regression component, which results in the highest rise in the resulting proxy--target correlation.

\begin{table}
\caption{Modeling PMOD using PSI data. The first ten components from the greedy search for regression components. Parameters $W$, $O,$ and component modes are selected without restrictions.}
\label{table008}
\begin{center}
\begin{tabular}{llccrrr}
N & Type & $W$ & $1/O$ & $R_c$ & $\Delta R_c$ \\
\hline
  1 & $E^{ }$ &     791.9 &  -  &       0.8604 & &  \\
  2 & $E^{+}$ &    4143 &     656.5 &       0.8742 &       0.0138 \\
  3 & $E^{ }$ &     342.1 &      27.22 &       0.8866 &       0.0124 \\
  4 & $E^{ }$ &      10.10 &      10.31 &       0.8930 &       0.0064 \\
  5 & $E^{-}$ &     431.3 &      26.81 &       0.8969 &       0.0039 \\
  6 & $E^{-}$ &       5.911 &       8.026 &       0.8995 &       0.0025 \\
  7 & $E_{d}^{-}$ &     119.0 &     114.1 &       0.9023 &       0.0028 \\
  8 & $E^{-}$ &    2305 &    1670 &       0.9041 &       0.0018 \\
  9 & $E_{d}^{-}$ &    4990 &     295.2 &       0.9054 &       0.0013 \\
 10 & $E_{d}^{+}$ &     139.2 &       8.544 &       0.9063 &       0.0009 \\
\hline
\end{tabular}
\end{center}
\end{table}
\begin{table}
\caption{Modeling ACRIM using PSI data. The first 16 components of the
  greedy search for regression components. Parameters $W$, $O,$ and component modes are
  selected without restrictions.}
\begin{center}
\begin{tabular}{llccrrr}
N & Type & $W$ & $1/O$ & $R_c$ & $\Delta R_c$ \\
\hline
  1 & $E_{d}^{+}$ &    5170 &    4420 &       0.7820 & &  \\
  2 & $E_{d}^{+}$ &    2519 &    1630 &       0.8342 &       0.0522 \\
  3 & $E^{+}$ &    1504 &     289.7 &       0.8538 &       0.0197 \\
  4 & $E^{-}$ &    3407 &     148.8 &       0.8771 &       0.0233 \\
  5 & $E^{ }$ &     362.4 &      27.34 &       0.8858 &       0.0087 \\
  6 & $E_{d}^{+}$ &     430.6 &    2183 &       0.8917 &       0.0059 \\
  7 & $E^{-}$ &    2340 &     300.9 &       0.8979 &       0.0062 \\
  8 & $E^{-}$ &     165.1 &      93.27 &       0.9021 &       0.0042 \\
  9 & $E^{ }$ &       9.638 &      10.60 &       0.9057 &       0.0035 \\
 10 & $E^{+}$ &    2571 &      27.38 &       0.9079 &       0.0022 \\
 11 & $E_{d}^{+}$ &      15.52 &      34.88 &       0.9094 &       0.0015 \\
 12 & $E^{-}$ &    1931 &     289.6 &       0.9111 &       0.0017 \\
 13 & $E_{d}^{+}$ &     124.5 &     112.1 &       0.9122 &       0.0011 \\
 14 & $E_{d}^{+}$ &      95.27 &     105.4 &       0.9151 &       0.0030 \\
 15 & $E_{d}^{+}$ &     153.4 &     109.7 &       0.9167 &       0.0015 \\
 16 & $E^{+}$ &     887.0 &       9.470 &       0.9176 &       0.0009 \\
\hline
\end{tabular}
\label{table009}
\end{center}
\end{table}

The results of this greedy search for the input data sets PSI and PMOD are
listed in Table~\ref{table008}; for comparison we also include
those for the PSI--ACRIM
pair (Table~\ref{table009}).  In these models we allowed the
parameters $W$ and $O$ to vary and also permitted envelope modes (upper and
lower) with both smoothed and detrended variants. In the last two columns
of the tables we list the ever-rising correlation values and correlation increments for the regression
solutions. Very different components
from the full library are involved.  Straightforward modeling of the PMOD (or ACRIM therefore) curve using
transformed variants of the PSI curve as components can clearly
be performed
formally with high precision. In principle, the set of components can
still be enlarged and the correlation level can be increased. However, the actual
benefits of a further refinement are quite minor. Most of the actual
predictive power is always concentrated in the very first components.
 
Unfortunately, actual hindcasts of these models are quite problematic.  
In Fig.~\ref{fig013} the two PSI-based hindcasts are plotted for the targets PMOD and ACRIM. The absolute
values of these curves depend on calibration, and we do not make an
effort to match them. 
We also overplot the backbones of the target curves to
illustrate that our predicting scheme tends to match the low-frequency
peculiarities.

The plots show that the hindcast
values are very different and show elements of strong nonstationarity. 

Consequently, we need to select components with care. Some of them
tend to fit into contingent features of the observations, and the
others model stationary aspects of the variability.

\begin{figure} 
\centerline{\includegraphics[width=0.5\textwidth,clip=]{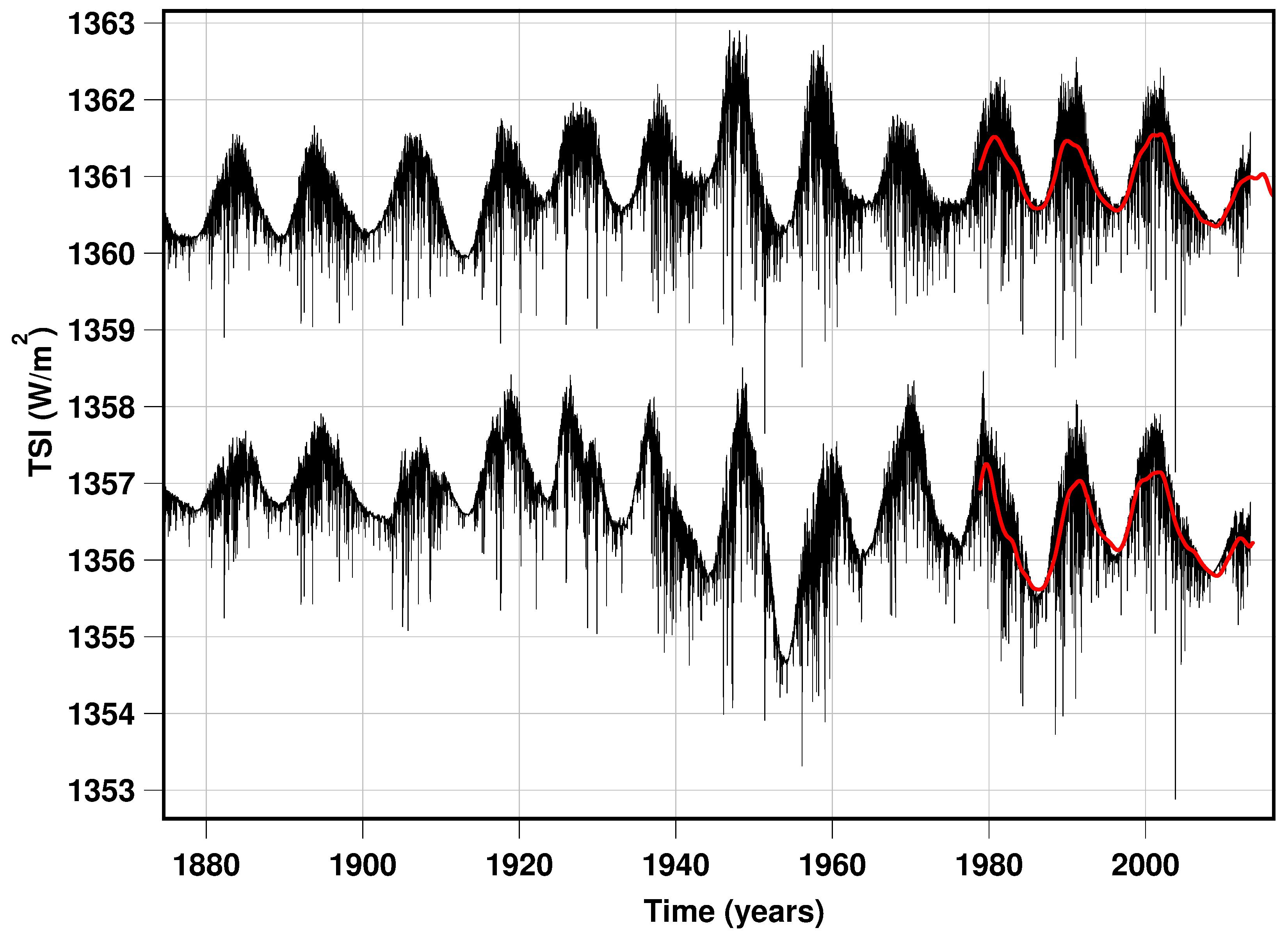}}
\caption{Hindcasting based on unrestricted multicomponent models. 
Using PSI to PMOD mapping (upper curve), PSI to ACRIM  (lower curve, shifted down by five units). In red we plot the low-frequency
backbones of the target curves.
The cross correlation between the two curves is only $R_c = 0.512$ (see Fig.~\ref{fig010}b).
} 
\label{fig013}
\end{figure} 

\section{Mutual correlations of the computed hindcasts for all proxy-target pairs.}\label{appb}

The different solutions for the hindcasting problem can be compared by computing their correlations
for the time intervals they share. There are 36 different correlation values for both
prediction methods (simplest and restricted multicomponent models).
We present the full list of correlations in Tables~\ref{table17} (simplest) and \ref{table18} (multicomponent).
Some solutions correlate at a very high level. For instance, the PSI to PMOD and PSI to RMIB 
predictions that are computed
using the simple method correlate at the level of $0.9997$. The worst case is between the SN to PMOD and SA to ACRIM predictions that are 
computed using multicomponent method ($R_c=0.8041$). Here the scatter between different solutions is a result of many contributing factors:
differences between PMOD and ACRIM (the main hindrance for better hindcasting), 
information content difference between proxies (SN against SA), and 
inherent modeling errors of the method.
These tables clearly show that the correlations on the diagonals of the subtables 
are more narrowly spread than the overall scatter (these are the pairs with common proxies). 
For the simple method, the spread on diagonals is $0.9957-0.9997,$ and for the full scale method it is $0.8879-0.9836$.  

\begin{table*}
\caption{Correlations between proxy-target combinations for the simplest model.}
\begin{tabular}{l|l|rrr|rrr|rrr|}
\multicolumn{2}{r}{} &
\multicolumn{3}{c}{PMOD} &
\multicolumn{3}{c}{RMIB} &
\multicolumn{3}{c}{ACRIM} \\
\cline{3-11}
\multicolumn{2}{c|}{} &  PSI &        SA &         SN &         PSI &        SA &         SN &         PSI &        SA &         SN \\  
\cline{2-11}
\multirow{3}{*}{PMOD}
 & PSI &     1.0000 &     0.9545 &     0.8832 &    {\bf 0.9997} &    0.9546 &     0.8839 &     0.9984 &     0.9522 &     0.8738 \\ 
 & SA &     0.9545 &     1.0000 &     0.9092 &     0.9519 &     {\bf 0.9993} &    0.9090 &     0.9580 &     0.9981 &     0.9008 \\ 
 & SN &     0.8832 &     0.9092 &     1.0000 &     0.8757 &     0.8980 &     {\bf 0.9996} &    0.8978 &     0.9175 &     \iti{0.9958} \\ 
\cline{2-11}
\multirow{3}{*}{RMIB}
 & PSI &     {\bf 0.9997} &    0.9519 &     0.8757 &     1.0000 &     0.9528 &     0.8763 &     \iti{0.9967} &    0.9484 &     0.8650 \\ 
 & SA &     0.9546 &     {\bf 0.9993} &    0.8980 &     0.9528 &     1.0000 &     0.8979 &     0.9562 &     \iti{0.9957} &    0.8880 \\ 
 & SN &    0.8839 &     0.9090 &     {\bf 0.9996} &    0.8763 &     0.8979 &     1.0000 &     0.8989 &     0.9181 &     0.9972 \\ 
\cline{2-11}
\multirow{3}{*}{ACRIM}
 & PSI &    0.9984 &     0.9580 &     0.8978 &     \iti{0.9967} &    0.9562 &     0.8989 &     1.0000 &     0.9589 &     0.8920 \\ 
 & SA &    0.9522 &     0.9981 &     0.9175 &     0.9484 &     \iti{0.9957} &    0.9181 &     0.9589 &     1.0000 &     0.9135 \\ 
 & SN &    0.8738 &     0.9008 &     \iti{0.9958} &     0.8650 &     0.8880 &     0.9972 &     0.8920 &     0.9135 &     1.0000 \\ 
\cline{2-11}
\end{tabular}
\tablefoot{The correlation maxima for the proxy-target pairs with common proxy are in 
listed in boldface and corresponding minima in italics. The full spread 
of correlations over all pairs is $0.8650-0.9997$.}
\label{table17}
\end{table*}

\begin{table*}
\caption{Correlations between proxy-target combinations for the multicomponent model.}
\begin{tabular}{l|l|rrr|rrr|rrr|}
\multicolumn{2}{r}{} &
\multicolumn{3}{c}{PMOD} &
\multicolumn{3}{c}{RMIB} &
\multicolumn{3}{c}{ACRIM} \\
\cline{3-11}
\multicolumn{2}{c|}{} & PSI & SA & SN & PSI & SA & SN & PSI & SA & SN \\  
\cline{2-11}
\multirow{3}{*}{PMOD}
 & PSI &    1.0000 &     0.9244 &     0.8358 &     {\bf 0.9836} &    0.8991 &     0.8465 &     0.9409 &     0.8746 &     0.8229 \\ 
 & SA  &    0.9244 &     1.0000 &     0.8677 &     0.9249 &     {\bf 0.9454} &    0.8675 &     0.8918 &     0.8975 &     0.8455 \\ 
 & SN  &    0.8358 &     0.8677 &     1.0000 &     0.8296 &     0.8310 &     0.9330 &     0.8151 &     0.8041 &     \iti{0.9128} \\ 
\cline{2-11}
\multirow{3}{*}{RMIB}
 & PSI &    {\bf 0.9836} &    0.9249 &     0.8296 &     1.0000 &     0.9017 &     0.8436 &     \iti{0.9340} &    0.8705 &     0.8181 \\ 
 & SA  &    0.8991 &     {\bf 0.9454} &    0.8310 &     0.9017 &     1.0000 &     0.8397 &     0.8804 &     \iti{0.8879} &     0.8147 \\ 
 & SN  &    0.8465 &     0.8675 &     0.9330 &     0.8436 &     0.8397 &     1.0000 &     0.8260 &     0.8190 &     {\bf 0.9376} \\ 
\cline{2-11}
\multirow{3}{*}{ACRIM}
 & PSI &    0.9409 &     0.8918 &     0.8151 &     \iti{0.9340} &    0.8804 &     0.8260 &     1.0000 &     0.8853 &     0.8122 \\ 
 & SA  &    0.8746 &     0.8975 &     0.8041 &     0.8705 &     \iti{0.8879} &    0.8190 &     0.8853 &     1.0000 &     0.8108 \\ 
 & SN  &    0.8229 &     0.8455 &     \iti{0.9128} &     0.8181 &     0.8147 &     {\bf 0.9376} &    0.8122 &     0.8108 &     1.0000 \\ 
\cline{2-11}
\end{tabular}
\tablefoot{See Table~\ref{table17} above for the notation. The full spread over all pairs is $0.8041-0.9836$.}
\label{table18}
\end{table*}

\end{document}